\documentclass[fleqn,usenatbib]{mnras}

\usepackage{newtxtext,newtxmath}

\usepackage[T1]{fontenc}
\usepackage{ae,aecompl}

\usepackage{graphicx}	
\usepackage{amsmath}	
\usepackage[svgnames]{xcolor}

\newcommand{\Rey}[1]{\overline{#1}}
\newcommand{\Fav}[1]{\widetilde{#1}}

\title{3D simulations of convective shell Neon-burning in a massive star}

\author[Georgy et al.]{Georgy, C.$^{1}$\thanks{E-mail: cyril.georgy@unige.ch}, Rizzuti, F.$^{2}$, Hirschi, R.$^{2,3}$, Varma, V.$^{2}$, Arnett, W. D.$^{4}$, Meakin C.$^{5}$, Mocak M.$^{6}$, Murphy A. StJ.$^{7}$, 
\newauthor
{Rauscher, T.$^{8,9}$}
\\
$^{1}$Département d'astronomie, Université de Genève, Chemin Pegasi 51, CH-1290 Versoix, Switzerland\\
$^{2}$Astrophysics Group, Lennard-Jones Laboratories, Keele University, Keele ST5 5BG, UK\\
$^{3}$Kavli IPMU (WPI), University of Tokyo, 5-1-5 Kashiwanoha, Kashiwa 277-8583, Japan\\
$^{4}$Steward Observatory, University of Arizona, 933 N. Cherry Avenue, Tucson AZ 85721, USA\\
$^{5}$Pasadena Consulting Group, 1075 N Mar Vista Ave, Pasadena, CA 91104 USA\\
$^{6}$NESS KE s.r.o, Továrenská 8, 04001 Košice, Slovakia\\
$^{7}$School of Physics and Astronomy, University of Edinburgh, Edinburgh EH9 3FD, UK\\
$^{8}$Department of Physics, University of Basel, 4056 Basel, Switzerland\\
$^{9}$Centre for Astrophysics Research, University of Hertfordshire, Hatfield AL10 9AB, United Kingdom}
\date{Accepted XXX. Received YYY; in original form ZZZ}

\pubyear{2024}

\begin{document}
\label{firstpage}
\pagerange{\pageref{firstpage}--\pageref{lastpage}}
\maketitle

\begin{abstract}
The treatment of convection remains a major weakness in the modelling of stellar evolution with one-dimensional (1D) codes. The ever increasing computing power makes now possible to simulate in 3D part of a star for a fraction of its life, allowing us to study the full complexity of convective zones with hydrodynamics codes. Here, we performed state-of-the-art hydrodynamics simulations of turbulence in a neon-burning convective zone, during the late stage of the life of a massive star. We produced a set of simulations varying the resolution of the computing domain (from $128^3$ to $1024^3$ cells) and the efficiency of the nuclear reactions (by boosting the energy generation rate from nominal to a factor of 1000). We analysed our results by the mean of Fourier transform of the velocity field, and mean-field decomposition of the various transport equations. Our results are in line with previous studies, showing that the behaviour of the bulk of the convective zone is already well captured at a relatively low resolution ($256^3$), while the details of the convective boundaries require higher resolutions. The different boosting factors used show how various quantities (velocity, buoyancy, abundances, abundance variances) depend on the energy generation rate. We found that for low boosting factors, convective zones are well mixed, validating the approach usually used in 1D stellar evolution codes. However, when nuclear burning and turbulent transport occur on the same timescale, a more sophisticated treatment would be needed. This is typically the case when shell mergers occur.
\end{abstract}




\section{Introduction}

To study the physics and evolution of stars, the most efficient and complete software tools available are one-dimensional (1D) stellar evolution models \citep{Heger2000a,2011ApJS..192....3P,2012A&A...537A.146E}. Not only can these models be employed to study the physics of stars, but they also represent a key aspect for interpreting stellar observations. Without an up-to-date and reliable grid of 1D stellar models, it would not be possible to obtain important information e.g. from asteroseismic measurements \citep[e.g][]{Aerts2003a}, or for isochrone fitting and age determination \citep[e.g.][]{Jorgensen2005a,Bossini2019a}. Despite the great progress made in the recent years to improve stellar evolutionary models, several uncertainties still affect the outcome of the models, undermining the accuracy of the predictions. These uncertainties arise from the multi-physical and multi-dimensional processes that occur in stars, often included only through simplified prescriptions.  Among the most concerning uncertainties in stellar modelling are determining the extent of convective regions and the mixing that occurs near the boundaries (convective boundary mixing, CBM).

For decades, 1D models have implemented the so-called ``mixing length theory'' \citep[MLT,][]{1958ZA.....46..108B}, a purely one-dimensional and local treatment of convection that, while simple and easy to implement, fails to appreciate the multi-dimensionality of the problem. This reflects in the well-known problem of having to add ad-hoc mixing beyond the convective boundary, often referred to as ``overshoot''. Over the years, many prescriptions have been suggested and improved to account for an extension of the convective region \citep[e.g.][]{Zahn1991a,Canuto1991a,1996A&A...313..497F,2000A&A...360..952H,Gabriel2014a}, but their validation and calibration has been difficult due to the impossibility of directly measuring the size of convective regions in observed stars. Indirect information can be deduced from asteroseismology \citep{2021NatAs...5..715P}, the Hertzsprung-Russell diagram \citep{2014A&A...570L..13C}, and eclipsing binaries \citep{Claret2016a}, but these methods are still heavily based on 1D models for interpreting the observations.

In the last two decades, multi-dimensional hydrodynamic simulations of stellar interiors have been used to constrain and parametrize CBM \citep[e.g.][]{1996A&A...313..497F,2007ApJ...667..448M}. This approach is sometimes called `321D-guided', referring to the fact that 1D models are improved with the results obtained from multi-D models. In fact, multi-D simulations are started from initial conditions assumed from the 1D models they are trying to validate, therefore they are subjected to the uncertainties of the 1D models. In addition, 3D models are not able to reproduce the long time-scales typical of the stellar evolution, due to the excessive computing resources required, therefore their results need to be generalized into more extensive prescriptions.

\begin{figure*}
\centering
\footnotesize
\includegraphics[width=0.49\textwidth]{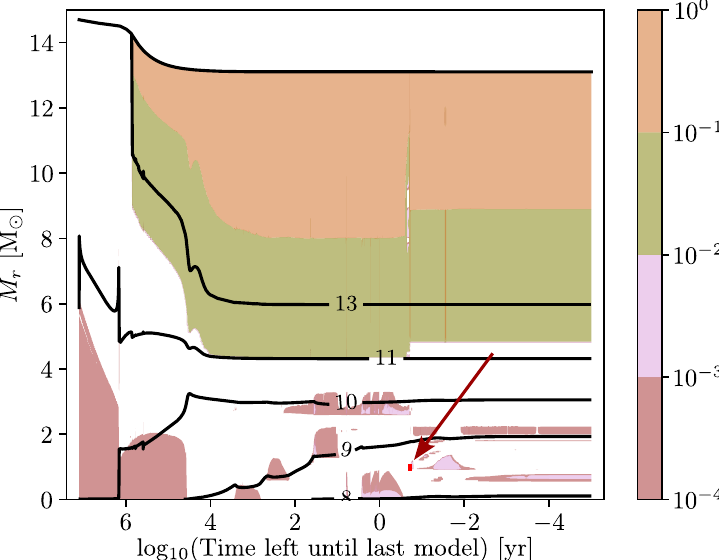}\hfill\includegraphics[width=0.49\textwidth]{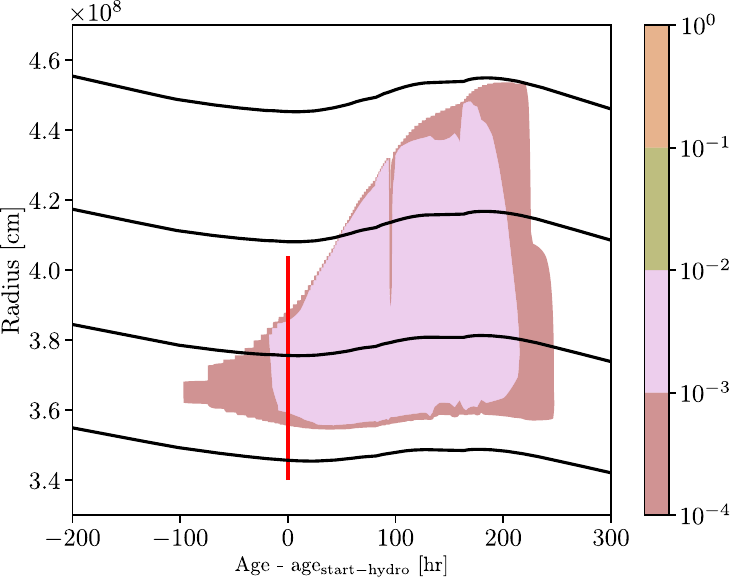}
\caption{\textit{Left}: Structure evolution (aka Kippenhahn) diagram of the 1D GENEC 15\,$M_\odot$ stellar model used as input for this study. The top solid black line shows the time evolution of the total mass. The other solid black lines are radial contours (values of $\log_{10}(r)$ in cm are indicated on the curves). The Mach number of the flow inside convective zones (shaded areas) is colour-coded. The extent of the computational domain covered by the 3D simulations corresponds to the red vertical bar (indicated by the red arrow). \textit{Right:}  Zoom-in on the neon shell convective region simulated in this study. $\text{age}_\text{start-hydro}$ corresponds to the time at which the 3D simulations start. Black solid lines are isomass contours ($M_r=0.9,1.0,1.1,1.2\,M_\odot$). These isomass contours show that the shell studied does not undergo any significant contraction or expansion during the Ne-burning phase. The 3D simulations presented in this paper cover a physical time of an hour or less, thus not even covering the width of the red vertical line in this figure. These diagrams are taken from \citet{Rizzuti2022b}.}\label{fig_kip}
\end{figure*}

Running multi-D hydrodynamic simulations of stars is very challenging due to the great amount of computing resources required. Consequently, an important limiting factor in the simulations is the Mach number, or the convective velocity of the fluid. The late burning phases of massive stars are normally characterised by large convective velocities, which make it easier and cheaper to perform hydrodynamic simulations of these environments using time-explicit methods. They normally include oxygen- and silicon-burning phases \citep{2007ApJ...667..448M,Arnett2009a,Viallet2013a,Couch2015a,2016ApJ...833..124M,2017MNRAS.465.2991J,2019ApJ...881...16Y}, and occasionally also neon-burning \citep{Rizzuti2022b,Rizzuti2023b}. Recently, these convective phases have also been explored with rotation \citep{Yoshida2021, McNeill2022, Fields2022}, magnetic fields \citep{Varma2021, Leidi2023} and both rotation and magnetic fields \citep{Varma2023}. Earlier phases, such as the main-sequence burning, helium- and carbon-burning, are characterised by slow convective velocities, which require large boosting in luminosity to make the hydrodynamic computation affordable \citep{2013ApJ...773..137G,Woodward2015a,2017MNRAS.471..279C,2021A&A...653A..55H,Herwig2023a,Andrassy2024a} or in two dimensions \citep{Baraffe2023a}. However, it is still not clear whether such boosting also introduces additional effects to the physics of the star.

In this work, we present a large grid of hydrodynamic simulations reproducing a neon-burning shell in a 15 M$_\odot$ star; this grid explores a wide range of resolution and luminosity. Some of these simulations have been used in \citet{Rizzuti2022b} to study entrainment. Here, we perform a detailed analysis of the dynamics and nucleosynthesis and their interplay. The presence of a nominal-luminosity run allows us to validate the results normally obtained through luminosity extrapolation; having simulations with different boosting factors allows us to study how different quantities scale with the luminosity. We also perform a detailed mean-field decomposition of some key equations, to study the statistical properties of the fluid. Finally, an explicit network of isotopes has been included for realistically reproducing nuclear burning, allowing us to study the evolution of the abundances and their variance. The results we present here contribute to the general understanding of stellar evolution through our detailed analysis and critical comparison to the initial conditions.

We organize the paper as follows: In Section~\ref{Initial_Cond}, we present the initial conditions and a general overview of the hydrodynamic simulations. In Section~\ref{Flow}, we present the results concerning the dynamics and nucleosynthesis of the convective flow and we present the detailed mean-field analysis performed with the Reynolds-averaging framework. In Section~\ref{nuclear}, we analyse the time evolution and transport of the chemical composition. Finally, we discuss our results and draw conclusions in Section~\ref{Conclu}.

\section{Initial conditions and overview of 3D Hydrodynamic simulations}\label{Initial_Cond}

In this paper, we simulate a neon burning shell of a 15\,$M_\odot$ star. The initial conditions for our 3D models are mapped from a 1D GENEC stellar evolution model run at solar metallicity ($Z=0.014$) and with the physical ingredients described in \citet{2012A&A...537A.146E}. The Schwarzschild criterion is used, and penetrative overshoot is included for core hydrogen and core helium burning phases only, with an overshooting distance $\ell_{\text{over}}=0.1\,H_P$. From carbon burning onward, an $\alpha$-chain network is used \citep[see][for details]{2004A&A...425..649H}. This input stellar model has been described in more detail in \citet{Rizzuti2022b}. The evolution of the structure of the model is presented in Fig.\ref{fig_kip} (\textit{left}) with a close-up of the neon-burning shell used as initial conditions for the 3D simulations presented in Fig.\,\ref{fig_kip} (\textit{right}). Compared to other convective episodes, neon-burning convective zones are relatively small and short-lived with typical spatial extents of the order of 10$^{8-9}$\,cm and temporal extents of weeks to months. These timescales are still too computationally expensive to be simulated completely in 3D \citep[though see][]{Rizzuti2023b}, so in this study we opt instead to simulate a statistically significant number of convective turnovers. 

\begin{figure}
\centering
\footnotesize
\includegraphics[width=0.5\textwidth]{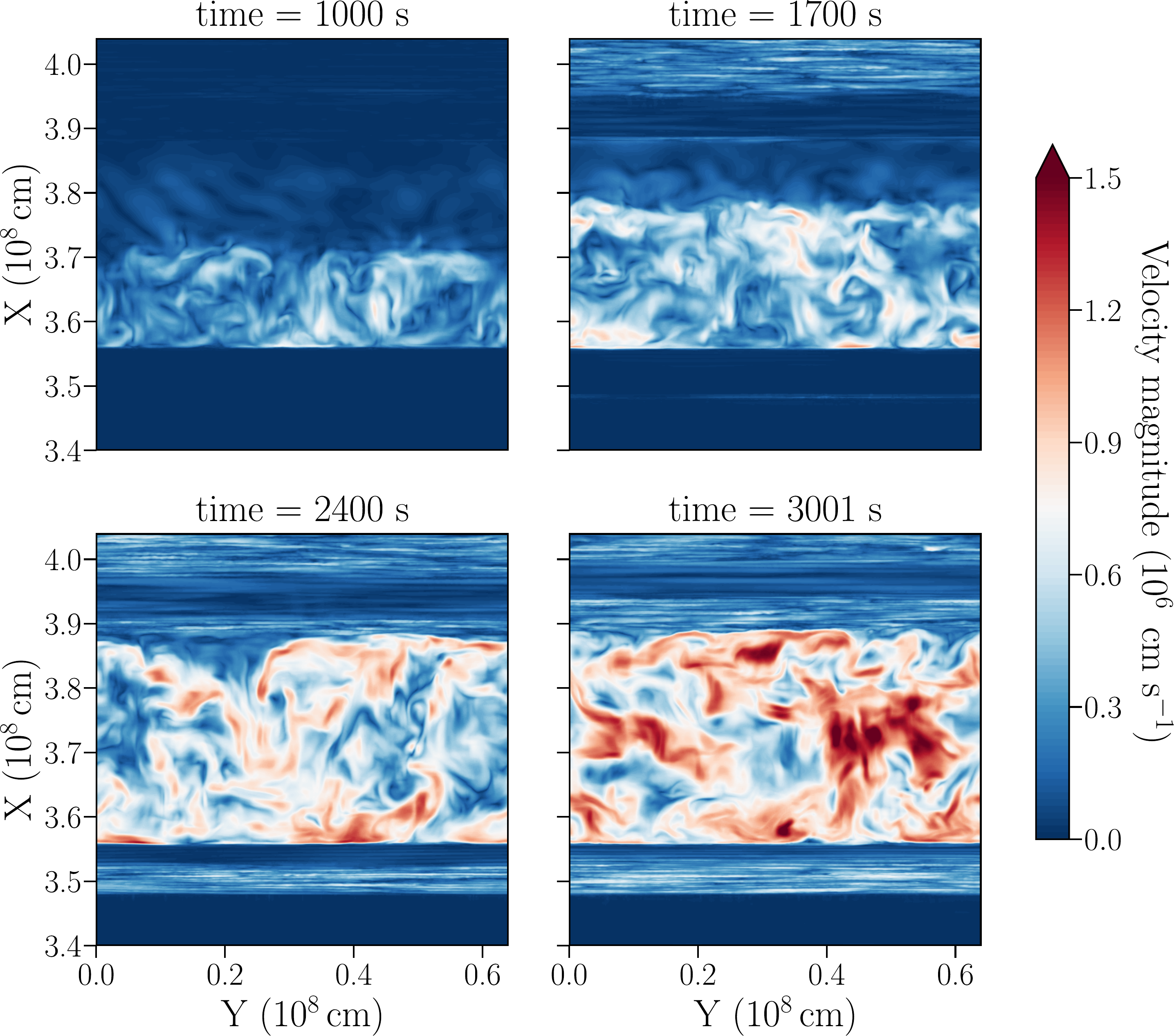}
\caption{Vertical cross sections of the velocity magnitude (values in colour scale) for the \texttt{Ex1} model, taken at 1000, 1700, 2400 and 3001 (last timestep) seconds through the simulation. The progression shows an increase in the velocity magnitude of fluid elements. It is possible to see in the upper panels the formation of the convective cell, and in the lower ones its growth into the upper stable region as expected from turbulent entrainment. Gravity waves are also visible in the stable region.}\label{fig_time_cross_section}
\end{figure}

Using the initial conditions described above, we produced a series of 3D hydrodynamic simulations using the \texttt{PROMPI} code \citep{2007ApJ...667..448M,2017MNRAS.471..279C}. We used a plane-parallel geometry and our computing domain is a cube of side equal to $0.65\times 10^8\,\text{cm}$ encompassing the convective shell, which occupies half of the domain, as well as stable (radiative) layers both below and above (see e.\,g. Fig.\,\ref{fig_time_cross_section}). The gravity is fixed according to the initial stellar model, and nuclear burning is followed with a minimal nuclear network of 4 species ($^{16}\text{O}$, $^{20}\text{Ne}$, $^{24}\text{Mg}$, and $^{28}\text{Si}$), $\alpha$-particles being considered to be at equilibrium \citep{Arnett1974a}. As already done in our previous works \citep{2017MNRAS.471..279C,Rizzuti2022b,Rizzuti2023b}, for a subset of our simulations, we multiplied the reaction rates and the neutrino generation rate by a so-called ``boosting'' factor taking the following values: $1$ (nominal case), $10$, $100$, and $1000$. This energy generation boosting provides two main advantages. First, the increased energy generation speeds up the nuclear burning and convection (convective velocities scale with the cubic root of the energy generation rate, see below). Higher convective velocities means that sufficient statistics (i.\,e. a reasonable number of convective turnovers) can be obtained with a smaller computing budget, as time steps are limited by the sound speed in \texttt{PROMPI} (time explicit integration). Second, and most important, having several simulations with different energy generation rates enables us to study the dependence of the results on the strength of turbulence. It is worth noting that this study includes a nominal case, i.e. exactly the same nuclear energy generation and neutrino losses as in the 1D input model. This enables a direct comparison to the 1D input model without the need for extrapolation.

\begin{table*}
\centering
\footnotesize
\caption{Summary of the models presented in this study, listed under names indicating their boosting factor and resolution. The properties are: resolution $N_\text{xyz}$, boosting factor of the nuclear energy generation rate $\epsilon$, physical time simulated $\tau_\text{sim}$ (s), global rms convective velocity $v_\text{rms}$ (cm s$^{-1}$), convective turnover time $\tau_\text{c}$ (s), time spent in quasi-steady state $\tau_\text{q}$ (s), number of convective turnovers simulated in quasi-steady state phase $n_c$, computational cost in CPU core-hours.}\label{tab_1}
\begin{tabular}{lcclcccccc}
&$N_\text{xyz}$&$\epsilon$&$\tau_\text{sim}$&$v_\text{rms}$&Ma&$\tau_\text{c}$&$\tau_\text{q}$&$n_c$&cost\\
 & & & [s] & [$10^6 \text{\,cm\,s}^{-1}$]&[$10^{-3}$] & [s] & [s] & & [$10^6$ hr]\\
\hline
\texttt{Ex10\_128}&$128^3$&$10$&1500&$1.58$&4.53&45&1250&27&0.04\\
\texttt{Ex10\_256}&$256^3$&$10$&1832&$1.70$&4.44&44&1600&36&0.36\\
\texttt{Ex1} $^a$&$512^3$&$1$&3037&$0.70$&1.93&100&700&7&11.4\\
\texttt{Ex10}&$512^3$&$10$&1004&$1.49$&3.96&46&700&15&4.66\\
\texttt{Ex100}&$512^3$&$100$&291&$3.72$&9.95&23&150&6&1.15\\
\texttt{Ex1000}&$512^3$&$1000$&73 $^b$&$8.00$&23.2&13&10&1&0.28\\
\texttt{Ex10\_1024}&$1024^3$&$10$&310 $^c$&$1.49$&3.94&47&310&6&48.2\\
\hline
\multicolumn{8}{l}{
\begin{minipage}[t][2.5em][t]{0.5\textwidth}
$^a$ Models with resolution $512^3$ are indicated by their boosting factor only, since they are the most studied in this work.
\end{minipage}
}\\
\multicolumn{8}{l}{
\begin{minipage}[t][2.5em][t]{0.5\textwidth}
$^b$ Time of the entire simulation, although the upper domain is reached at about 30 s, as it can be seen in Fig.~\ref{fig_STKE}.
\end{minipage}
}\\
\multicolumn{8}{l}{$^c$ Model \texttt{Ex10\_1024} was restarted from the \texttt{Ex10} simulation at 500 s.}\\ 
\end{tabular}
\end{table*}

We summarise the models presented in this study and their most important properties in Table~\ref{tab_1}. One can estimate an effective Reynolds number for our simulations as Re $\sim (n/2)^{4/3}$, where $n$ is the resolution, considering that the radial extent of the convective zone is half of the domain and thus covers $n/2$ cells \citep[see][]{Arnett2019}. This is a reasonable assumption, which is confirmed by the cross-sections of Figs. \ref{fig_time_cross_section} and \ref{fig_res_cross_section} during the quasi-steady state. In this way, we obtain effective Reynolds numbers of 256, 645, 1625 and 4096 for the resolutions $n=128$, 256, 512 and 1024, respectively. For large Reynolds numbers, we can assume the regime of turbulent cascade \citep{1932ZA......5..117B,Kolmogorov1941a}, where kinetic energy cascade is driven from the large scales. If we use $\text{Re} \ge 1000$ as the condition for the turbulent regime, we see that the highest resolutions $512^3$ (and $1024^3$) reach the numerically turbulent regime.  We thus focus our analysis on simulations at the $512^3$ resolution throughout the paper and shorten their code name to include only their boosting factor for convenience (\texttt{Ex1}, \texttt{Ex10}, \texttt{Ex100}, \texttt{Ex1000}). This being said, we analyse simulations with different resolutions to test the dependence of the results on the grid size in Section \ref{Flow}. Finally, we confirm from Table \ref{tab_1} that increasing the boosting factor by a factor of $10$ increases the convective velocity roughly by a factor of $10^{1/3}\simeq 2.15$ ($v_\text{rms}\sim\epsilon^{1/3}$, \citealt{Arnett2018b}, where $\epsilon$ is the energy generation rate). This in turn decreases the turnover, $\tau_c$, as expected. The general thermodynamics properties of the simulated region of the star remain mostly unchanged by changing the boosting factor, thus the sound speed also remains the same. The Mach number scales therefore with $v_\text{rms}$, and is small in any case : about  $2\cdot 10^{-3}$ for the \texttt{Ex1} case to bout $20\cdot 10^{-3}$ for the \texttt{Ex1000} case. We also confirm that simulations at different resolutions with the same boosting factor have approximately the same $v_\text{rms}$ , $\tau_c$, and $\text{Ma}$.

\section{General properties of the convective flow}\label{Flow}

In this section, we present our analysis of the convective flow with a particular emphasis on the velocity field and the turbulent kinetic energy.

\subsection{Flow velocity and specific kinetic energy}

Figure \ref{fig_time_cross_section} shows vertical cross sections of model \texttt{Ex1} taken at key phases of the simulation, with the velocity magnitude represented in colour scale. These cross sections allow us to see the time evolution of the velocity field in this simulation, which is representative of all our models. At the beginning of the simulation ($t=1000$ and 1700\,s), the nuclear burning at the bottom of the neon shell drives turbulent motions (initially plumes and later on eddies) which gradually fill the region that was convective in the 1D input stellar model. After an initial transient (the duration of which strongly depends on the boosting factor), convection is fully developed and the simulation enters a quasi-steady state (at around $2400\,\text{s}$ for the \texttt{Ex1} model). As time proceeds, the convective region grows and entrains material from the stable regions, as can be seen at the top boundary by comparing the cross sections at $t=2400$ and 3001\,s

\begin{figure}
\centering
\footnotesize
\includegraphics[width=.45\textwidth]{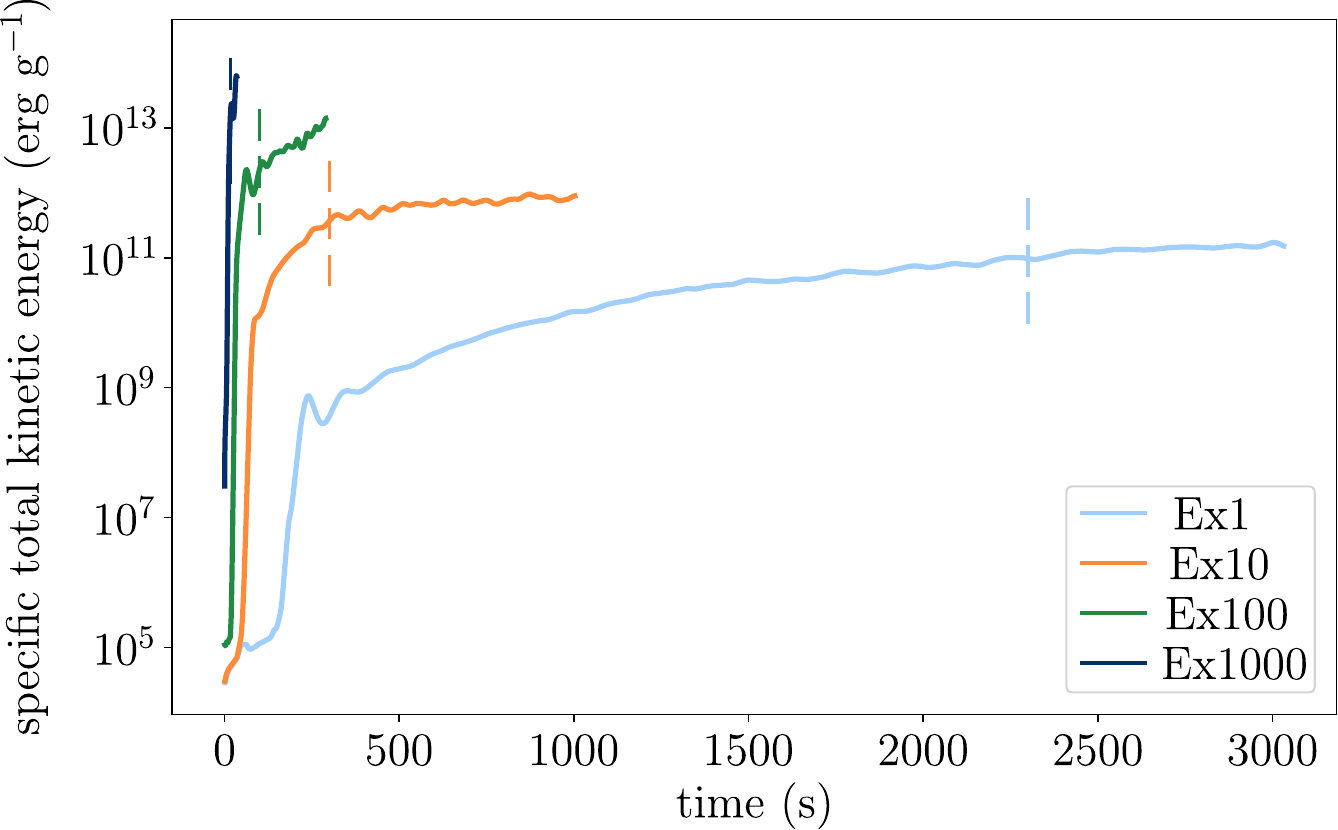}
\caption{Time evolution of the specific kinetic energy integrated over the computational domain, for the four models \texttt{Ex1}, \texttt{Ex10}, \texttt{Ex100}, \texttt{Ex1000}. The trends cover the entire simulated time range. After an initial transient, whose duration depends on the boosting, the simulations enter a quasi-steady state (starting time indicated by the vertical dashed lines), which lasts until the upper boundary of the domain is reached (\texttt{Ex100}, \texttt{Ex1000}) or the simulation is otherwise terminated (\texttt{Ex1}, \texttt{Ex10}). As expected, models with larger boosting factors reach higher kinetic energies.}\label{fig_STKE}
\end{figure}

Another way to follow the time evolution of the velocity in our simulations is via the specific total kinetic energy $(v_x^2+v_y^2+v_z^2)/2$  (see Fig.~\ref{fig_STKE}). The initial and sudden rise in kinetic energy characterises the initial transient. Its length depends on the boosting factor, being considerably longer for the non-boosted model (\texttt{Ex1}). After the transient, the simulations enter a quasi-steady state, where the average kinetic energy remains constant or slowly increases. This phase is also characterised by periodic pulses in kinetic energy of approximately the same timescale as the convective turnover time. These pulses are related to the time delay between the formation and rise of large-scale eddies in the convective zone and their dissipation \citep{2007ApJ...667..448M,Arnett2011a,Viallet2013a,Arnett2015a}. 

\begin{figure}
\includegraphics[width=\columnwidth]{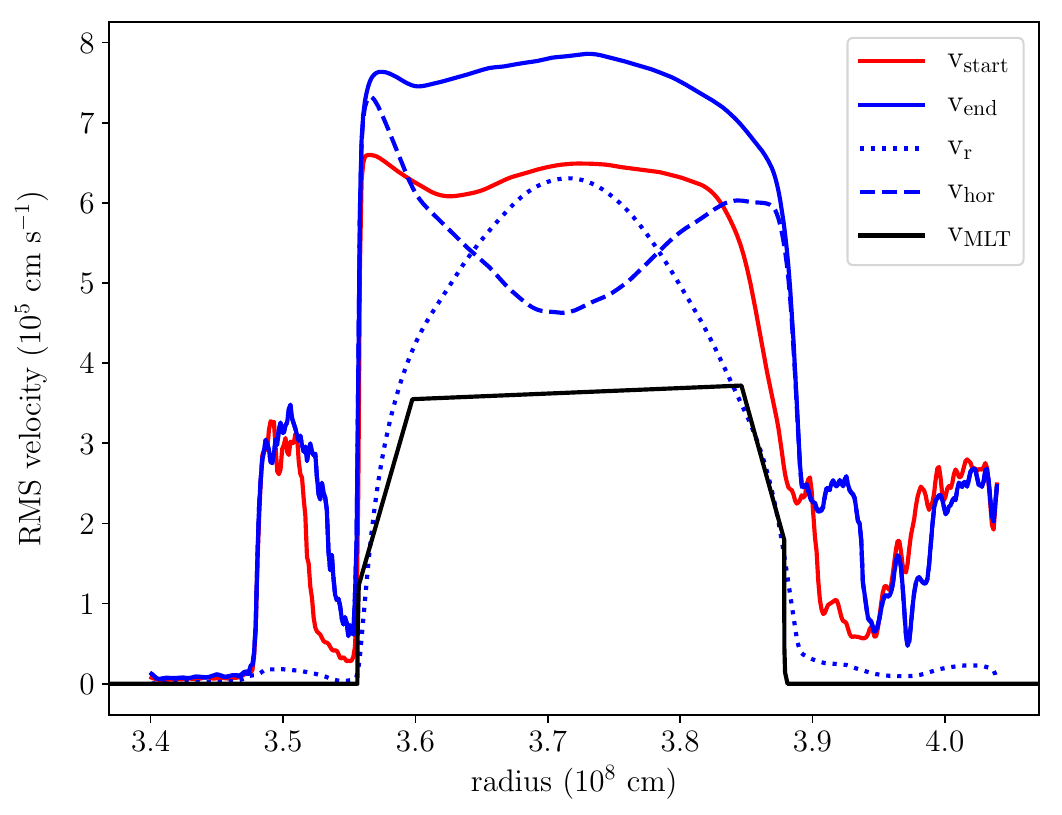}
\caption{Radial profiles of different velocity components. In black, the mixing-length-theory velocity v$_\text{MLT}$ of the 1D stellar model used as initial conditions for all simulations. In red, the root-mean-square velocity v$_\text{start}$ at the beginning of the quasi-steady state in \texttt{Ex1} simulation, averaged over one convective turnover. In blue, different components of the convective velocity at the end of \texttt{Ex1} simulation, averaged over one convective turnover: root-mean-square v$_\text{end}$ (solid), radial v$_\text{r}$ (dotted) and horizontal v$_\text{hor}$ (dashed) velocity components.}\label{fig_radial_profiles}
\end{figure}

The roughly constant or slow rise of convective velocities is confirmed by comparing the radial profiles of the root-mean-square velocity between the start (red solid line in Fig.~\ref{fig_radial_profiles} for \texttt{Ex1} model) and the end of the quasi-steady state phase (blue solid line). In addition to a mild increase in magnitude, the outward shifting of the upper boundary of the convective zone expected from entrainment is visible. This figure also confirms findings from previous 3D simulations \citep[see e.g.][]{2017MNRAS.471..279C,2017MNRAS.465.2991J, Mocak2018a}. The radial velocity component (dotted line) peaks in the centre of the convective zone while the horizontal component peaks at the boundaries. This reflects the convective motions and the u-turning of fluid elements at the boundaries. The non-negligible velocities outside the convective zone are produced by gravity waves. Furthermore, using our nominal case simulation (\texttt{Ex1}), we can compare convective velocities directly between 3D and 1D models. We see that convective velocities in our 3D simulations are about twice as large as the velocity predicted by mixing-length-theory in the 1D GENEC model (around $3.6\times10^5$ cm s$^{-1}$, black solid line) consistent with the results of \citet{2017MNRAS.465.2991J}.

\begin{figure}
\centering
\footnotesize
\includegraphics[width=0.5\textwidth]{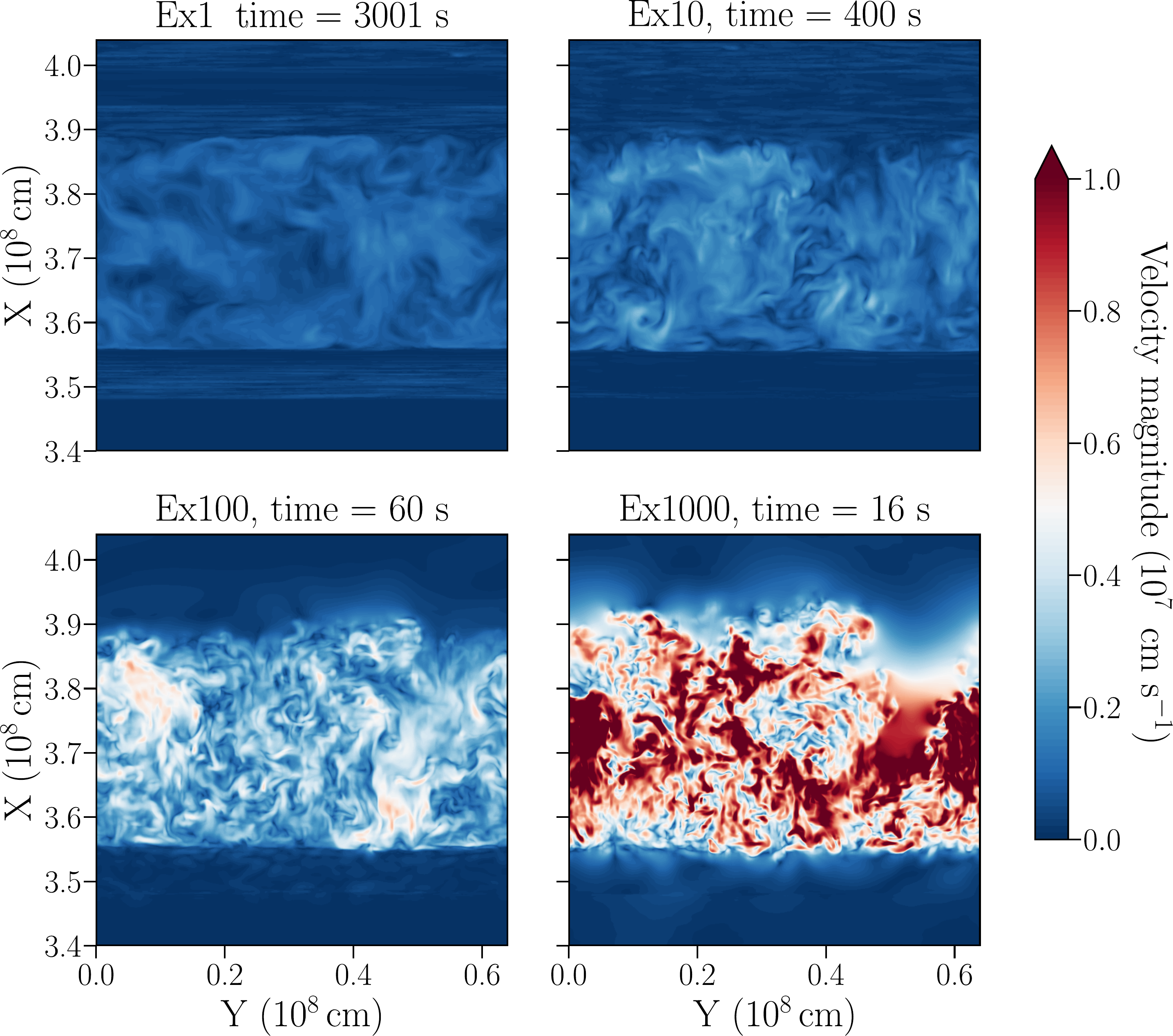}
\caption{Same as Fig. \ref{fig_time_cross_section}, but for the four models \texttt{Ex1}, \texttt{Ex10}, \texttt{Ex100}, \texttt{Ex1000} taken at 3001, 400, 60, 16 seconds, respectively. Timesteps were chosen so that the upper convective boundary is located at $3.89\times10^8$ cm in all simulations, for comparison. It is clear that one major effect of boosting the nuclear energy generation rate is an increase in the velocity magnitude of fluid elements, therefore in kinetic energy, as confirmed by Fig.~\ref{fig_STKE}.}\label{fig_boost_cross_section}
\end{figure}

The boosting of the nuclear energy generation rate has a strong impact on the evolution of our simulations. This is clearly visible in Fig.~\ref{fig_boost_cross_section}, where we compare the velocity fields between our four $512^3$ simulations with different boosting factors (see Table \ref{tab_1}). Since the time scale of the evolution of the models is affected by the boosting, as can be seen in Fig.~\ref{fig_STKE}, we choose here to compare the different simulations when their upper convective boundary has reached a radial location of approximately $3.89\times 10^8$ cm, which is the maximum extension of the convective zone in the non-boosted model. This corresponds to a time of 3001, 400, 60 and 16 seconds for the \texttt{Ex1}, \texttt{Ex10}, \texttt{Ex100} and \texttt{Ex1000} models, respectively (during the quasi-steady state phase of these simulations). Looking at the highest velocity in the snapshot shown in Fig.~\ref{fig_boost_cross_section} (coloured in red), we can see that larger boosting factors produced a higher kinetic energy of the fluid, which is confirmed in Fig.~\ref{fig_STKE}.

\begin{figure}
\centering
\footnotesize
\includegraphics[width=0.5\textwidth]{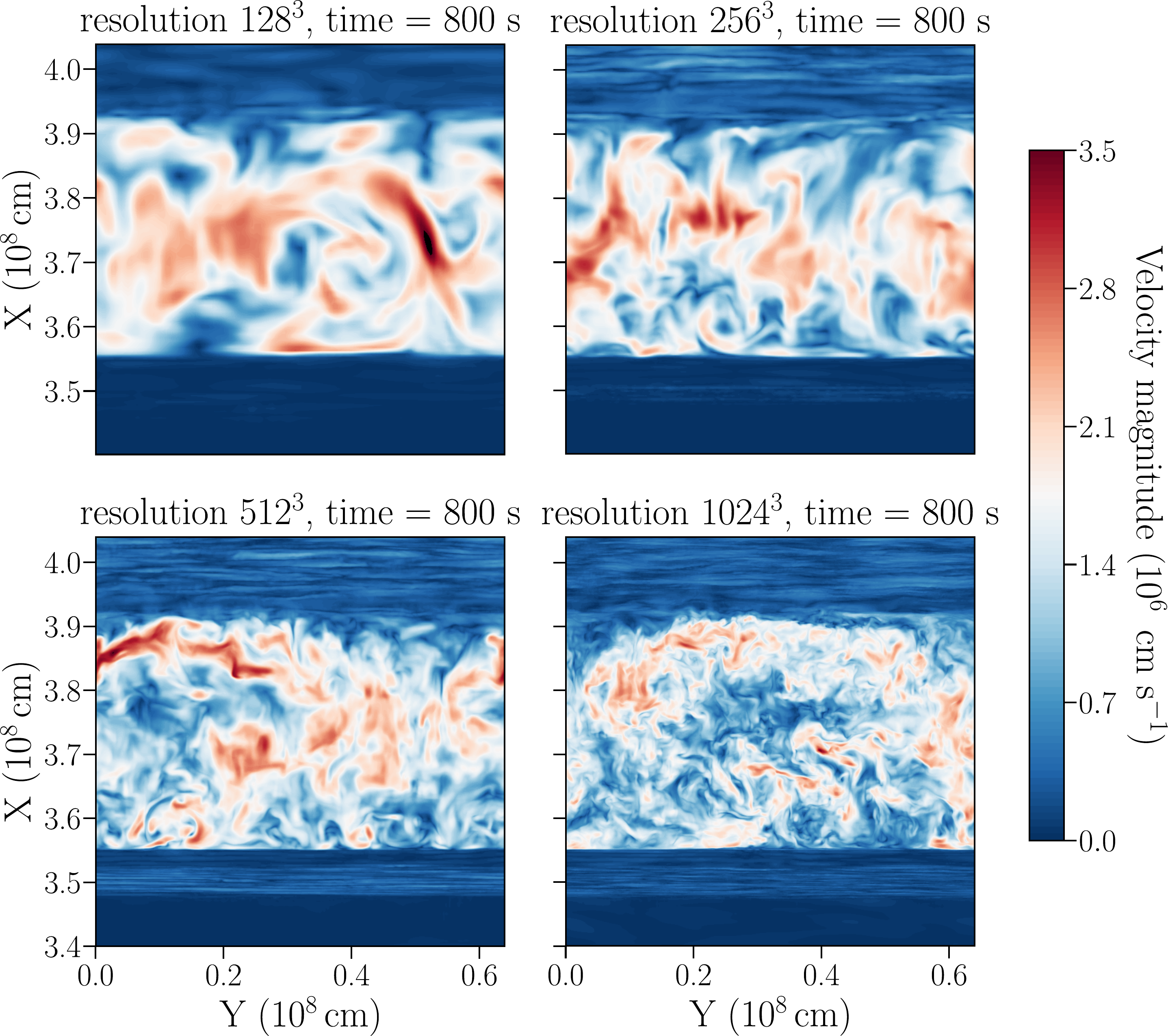}
\caption{Same as Fig. \ref{fig_time_cross_section}, but for the four different resolutions $128^3,256^3,512^3,1024^3$ with boosting factor 10 (see Table \ref{tab_1}), taken at 800 seconds from the beginning of each simulation. Since a higher resolution is linked to a smaller dissipation range, it is expected that the simulations predict eddies on a smaller scale when the resolution is increased.}\label{fig_res_cross_section}
\end{figure}

In order to test whether there is a dependence of the flow velocity on resolution, we present in Fig.~\ref{fig_res_cross_section} the velocity fields of four simulations with the same boosting factor and initial conditions, but different resolution. It is clear from the figure that the small scale features of convection depend on the mesh size we choose for our simulations. In particular, according to the ILES (implicit large eddy simulation) paradigm, the grid scale sets the limits for the numerical dissipation of kinetic energy, which mimics the effects of viscosity. For this reason, if we increase the resolution of our models the dissipation scale decreases, allowing the simulations to produce eddies on a smaller scale, which are closer to the real case scenario, as visible in Fig.~\ref{fig_res_cross_section} and discussed further in Sect.\,\ref{TKE_spectra}. 

\begin{figure}
\centering
\footnotesize
\includegraphics[width=.45\textwidth]{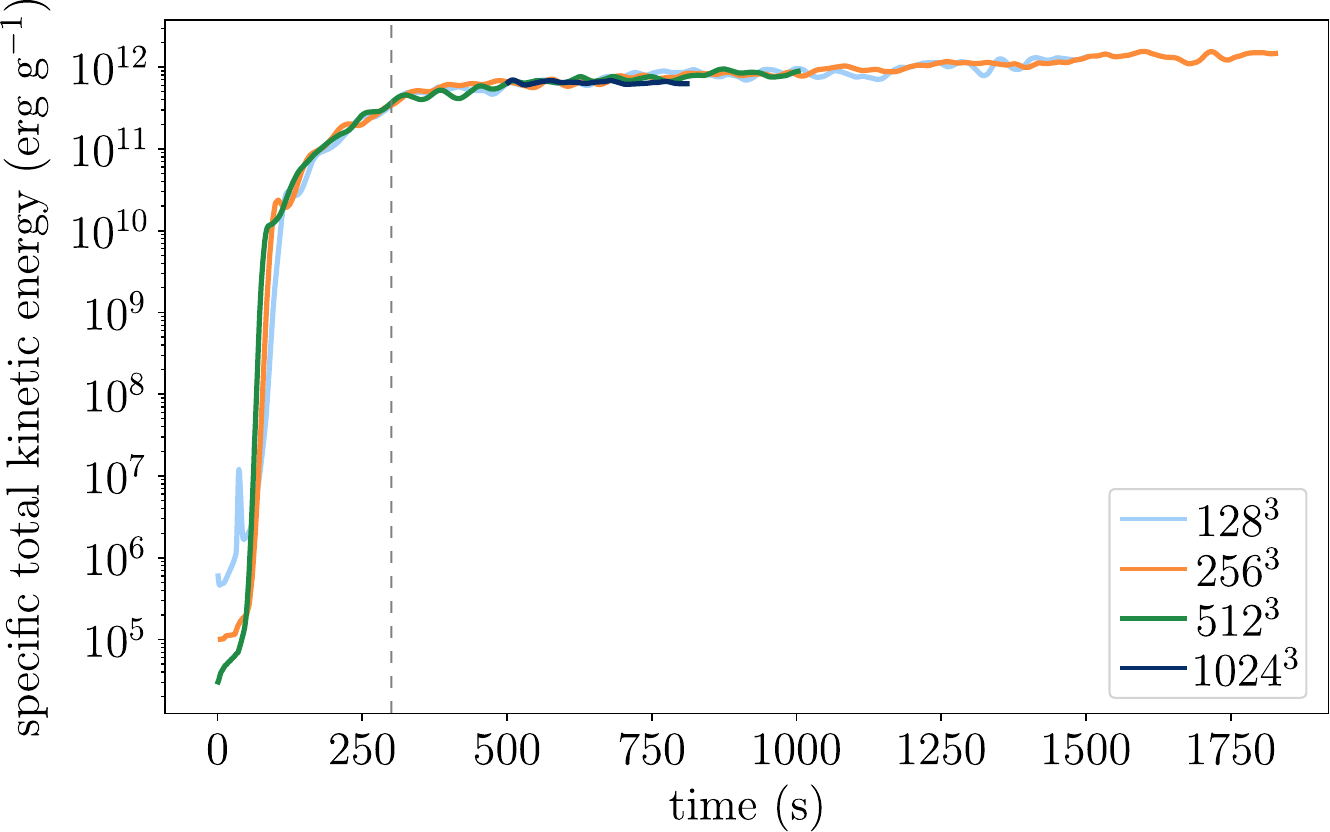}
\caption{Same as Fig. \ref{fig_STKE}, but for the four models with different resolutions $128^3,256^3,512^3,1024^3$ and same boosting factor 10. The simulations share a very similar evolution. The vertical dashed line is the beginning of the quasi-steady state. }\label{fig_STKE_res}
\end{figure}

At large scale, however, the structures look similar at different resolutions so we do not expect the bulk properties of the convective region to depend on resolution. This is confirmed in Fig.~\ref{fig_STKE_res} showing the time evolution of the specific total kinetic energy for the four simulations with different resolution and same boosting factor (\texttt{Ex10\_128}, \texttt{Ex10\_256}, \texttt{Ex10}, \texttt{Ex10\_1024}, see Table \ref{tab_1}). As expected, the simulations have a very similar evolution, in agreement with the fact that global properties (like $v_\text{rms}$ and $\tau_c$ estimated in  Table \ref{tab_1}) are comparable for all simulations with same boosting factor and thus do not depend on resolution. 

High resolution is nevertheless needed to better resolve convective boundaries as discussed in Sect.\,\ref{Mean_Field}.

\subsection{Turbulent kinetic energy spectra}\label{TKE_spectra}
A different approach to study the kinetic energy of the simulations is to compute the turbulent kinetic energy spectra. In order to do so, we performed 2D fast Fourier transforms of different velocity components on horizontal planes at constant height, within the convective zone. We also normalised the spectra by dividing them by $k^{-5/3}$, which is the power law scaling for the inertial range \citep{Kolmogorov1941a}. In this way, the regions where the spectra have a horizontal slope correspond to the inertial range \citep[see e.g.][and references therein]{2017MNRAS.471..279C}.

\begin{figure*}
\centering
\footnotesize
\includegraphics[width=0.9\textwidth]{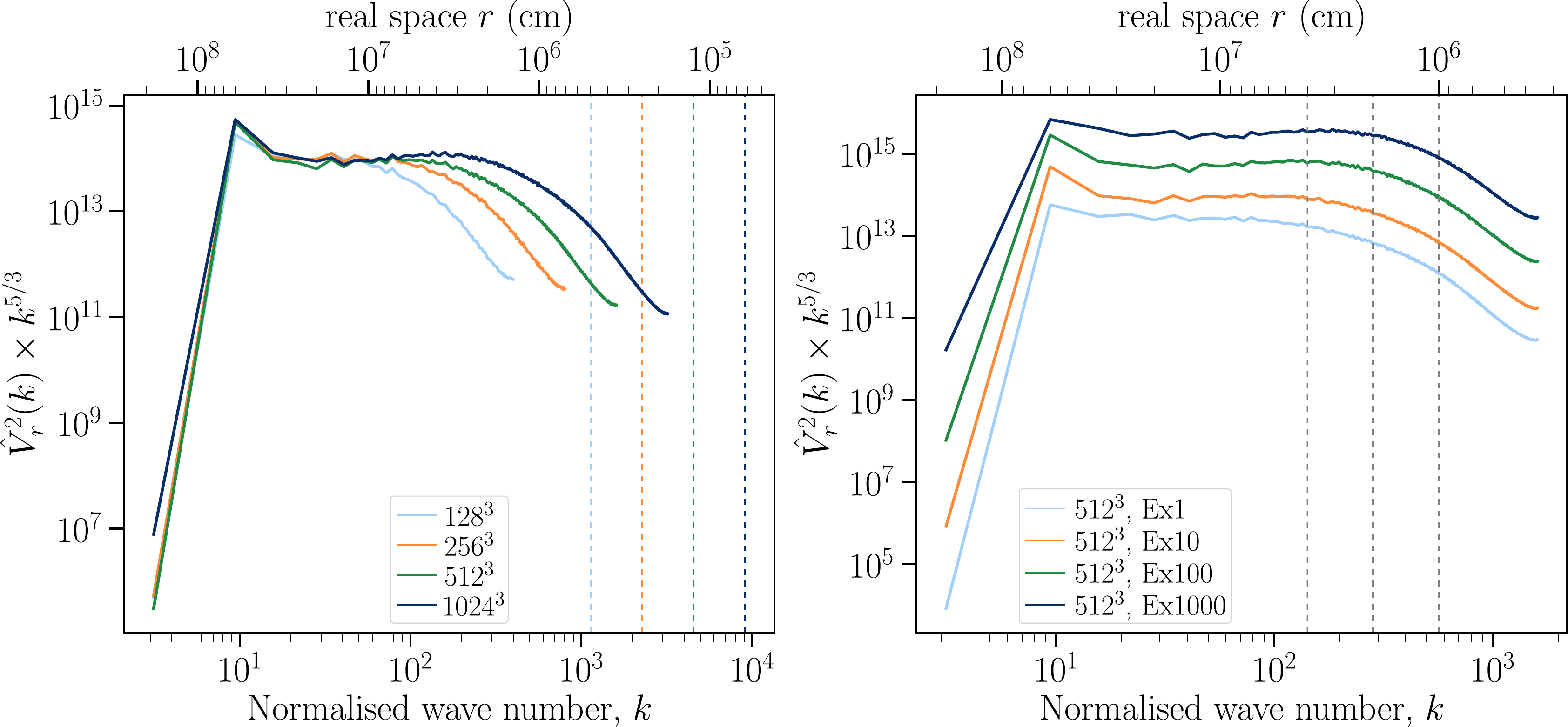}
\caption{Two-dimensional Fourier transforms of the radial velocity multiplied by $k^{5/3}$, for the four different resolutions $128^3,256^3,512^3,1024^3$ with a boosting factor 10 (left-hand panel), and for the four different boosting factors of the $512^3$ resolution models (right-hand panel), taken during the quasi-steady state phase, at the centre of the convective zone. The horizontal axes show both the Fourier space $k=\sqrt{k_y^2+k_z^2}$ (bottom axis) and the real space $r$ (top axis). The vertical dashed lines show in the left-hand plot the grid size 
for each resolution, and in the right-hand plot they show 32, 16, 8 times the grid size for the $512^3$ models.}\label{fig_res_spectra}
\end{figure*}
\begin{figure*}
\centering
\footnotesize
\includegraphics[width=\textwidth]{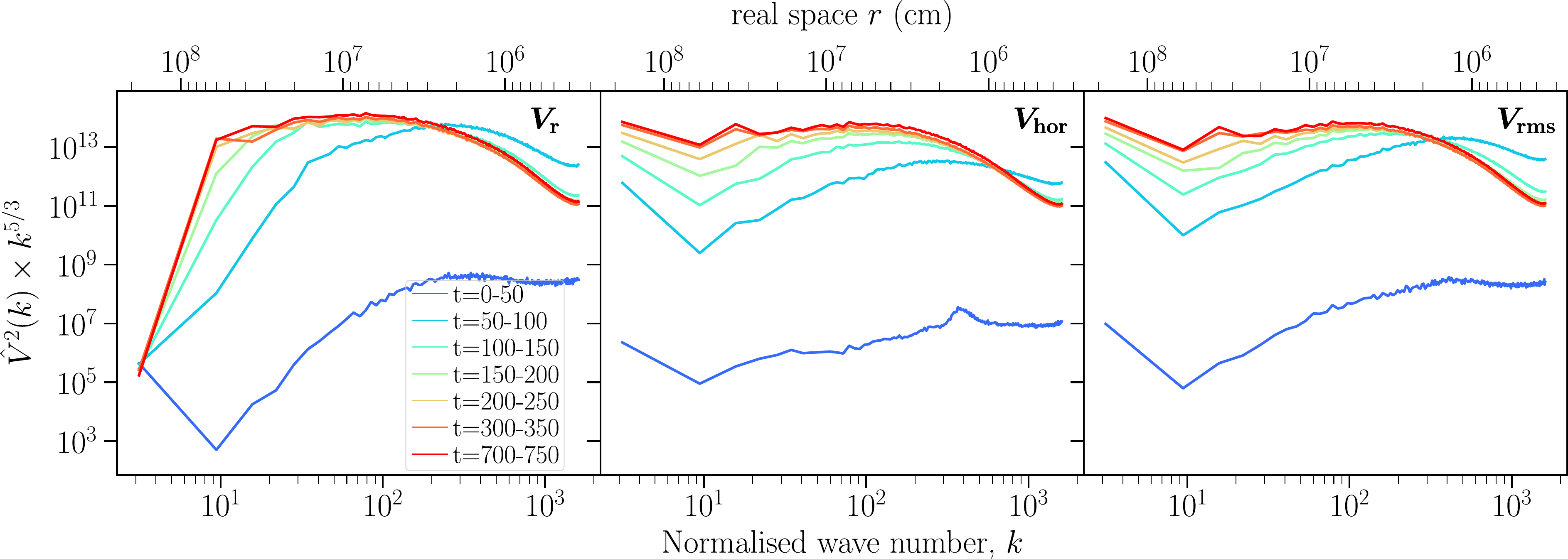}
\caption{Same as Fig. \ref{fig_res_spectra}, but for the Fourier transforms of the radial velocity (left-hand panel), horizontal velocity (central panel) and total root-mean-square (rms) velocity (right-hand panel), taken at different times through the \texttt{Ex10} simulation, with an averaging window of 50 seconds. It can be clearly seen the progression toward the quasi-steady state.}\label{fig_vel_spectra}
\end{figure*}
\begin{figure*}
\centering
\footnotesize
\includegraphics[width=\textwidth]{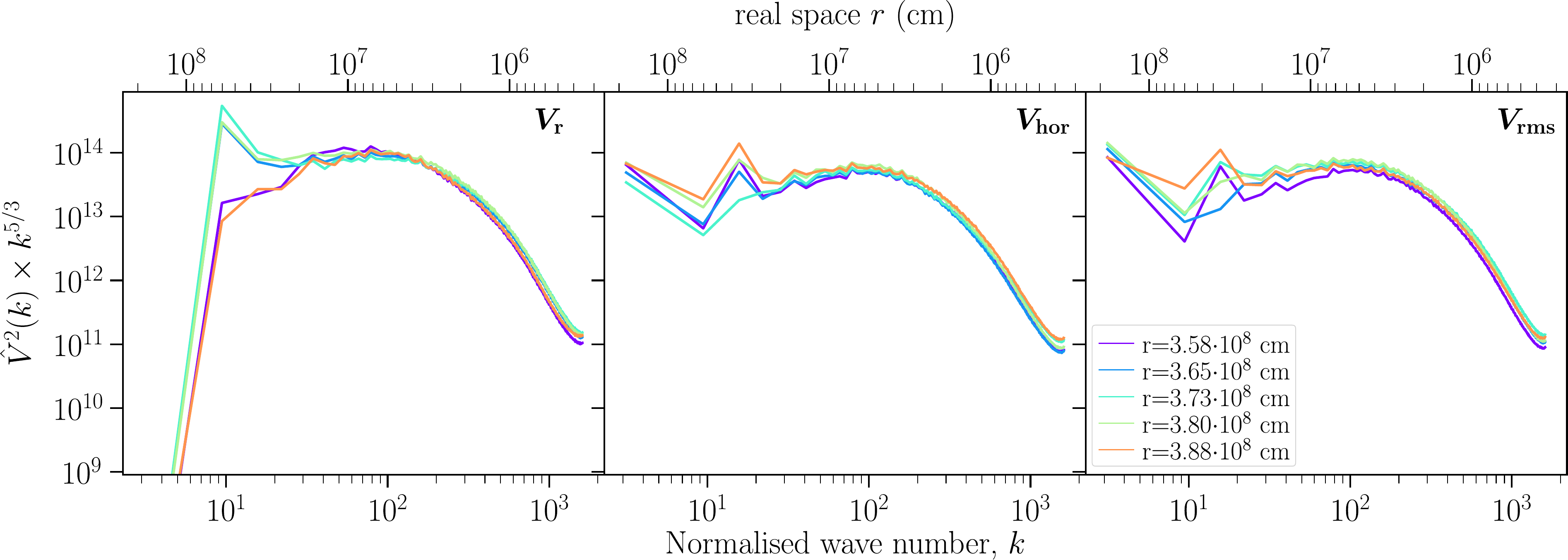}
\caption{Same as Fig. \ref{fig_vel_spectra}, but for the Fourier transforms taken during the quasi-steady state phase (500 seconds, \texttt{Ex10}) at different radial locations inside the convective zone.}\label{fig_rad_spectra}
\end{figure*}

We show in Fig.~\ref{fig_res_spectra} a comparison between the spectra of the radial velocity squared, for different resolutions (\textit{left panel}) and different boosting factors (\textit{right panel}). The spectra are averaged through the entire quasi-steady state phase of each simulation (see Fig.~\ref{fig_STKE}). As expected, increasing the resolution has the effect of extending the inertial range plateau toward higher $k$, because dissipation occurs at smaller scales $r$ in ILES. The vertical dashed lines indicate the grid size for each resolution. We can see that the dissipation range is a $5-10$ times larger than the grid size. On the other hand, if we increase the boosting factor while keeping the resolution fixed, we witness a rise in the velocity magnitude without changes in the length of the plateau, as visible in the right-hand panel of Fig.~\ref{fig_res_spectra}.

The spectra presented in Fig.~\ref{fig_res_spectra} are taken in the bulk of the convective region and during the quasi-steady state phase. It is interesting to find out if the spectra vary with location and time. In Fig.~\ref{fig_vel_spectra}, we study the time evolution of the velocity spectra for the \texttt{Ex10} model from the beginning of the simulation to the quasi-steady state, with time-averaging windows of 50 seconds, which is approximately the convective turnover time for this model (see Table~\ref{tab_1}). The spectra are taken near the bottom of the convective zone (at a radius of $3.58\times 10^8$ cm), and we present results for the radial velocity (\textit{left-hand side}), the horizontal velocity (\textit{central panel}) and the total root-mean-square velocity (\textit{right-hand side}). As can be seen from the plots, at the beginning of the simulation (during the initial transient), the spectra have a peak at high $k$, i.e. at small scales. This happens because convection is at a very early stage, and eddies on the smallest scales are dominant. As time passes, the velocity magnitude increases and the peaks are shifted toward smaller $k$ as the turbulent flow fills the entire region that was convective in the 1D input stellar model. After the initial transient (lasting about 300\,s), the spectra do not vary significantly and they assume the more familiar shape of Fig.~\ref{fig_res_spectra}, which is characteristic of homogeneous and isotropic turbulence.

In a similar way, we compare in Fig.~\ref{fig_rad_spectra} the spectra of the velocity and its components at different radial locations inside the convective zone, from $3.58$ to $3.88\times 10^8$ cm, during the quasi-steady state phase (500-550 seconds, \texttt{Ex10}). The spectra at different heights are very similar, which means that convection remains turbulent and generally isotropic throughout the convective zone. The main exception is the radial velocity at the lowest $k$ values (largest scales) near convective boundaries. Indeed, both the uppermost and lowermost spectra have a lower radial velocity magnitude around $k=10$ compared to the bulk of the convective region. This is easily explained if we consider that the convective boundaries are limiting the velocity in the radial direction, while the horizontal components of velocity are not affected by this restriction. Instead, they are stronger near the convective boundaries and present a peak around $k=11$ which is less important for central regions.

\subsection{Mean field (RA-ILES) analysis of the turbulent kinetic energy}\label{Mean_Field}

\begin{center}
\begin{figure*}
\includegraphics[width=.7\textwidth]{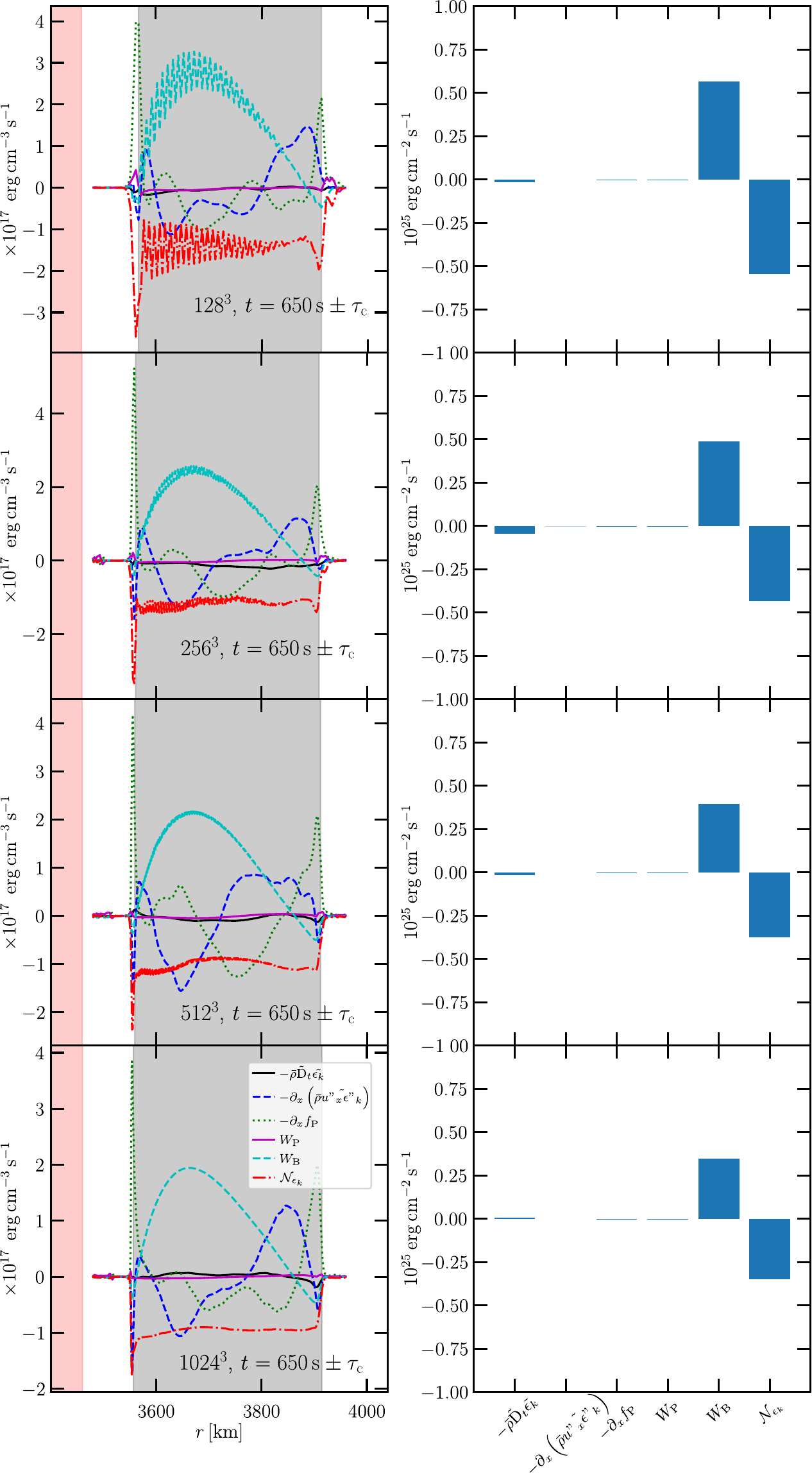}
\caption{\textit{Left column: }Mean kinetic energy equation terms as a function of radius for various resolutions. From top to bottom: $128^3$, $256^3$, $512^3$,$1024^3$. The time averaging is made over 2 turnover timescales centered on $t=650\,\text{s}$ for each resolution. The meaning of each curve is described in the caption in the first row. The grey area shows the convective region, and the red area the region where we applied a damping. \textit{Right column: }Radial integration of each terms of the mean kinetic equation for the same resolutions as in the left column.}
\label{fig_RANS_MKEE_reso}
\end{figure*}
\end{center}

\begin{center}
\begin{figure*}
\includegraphics[width=.65\textwidth]{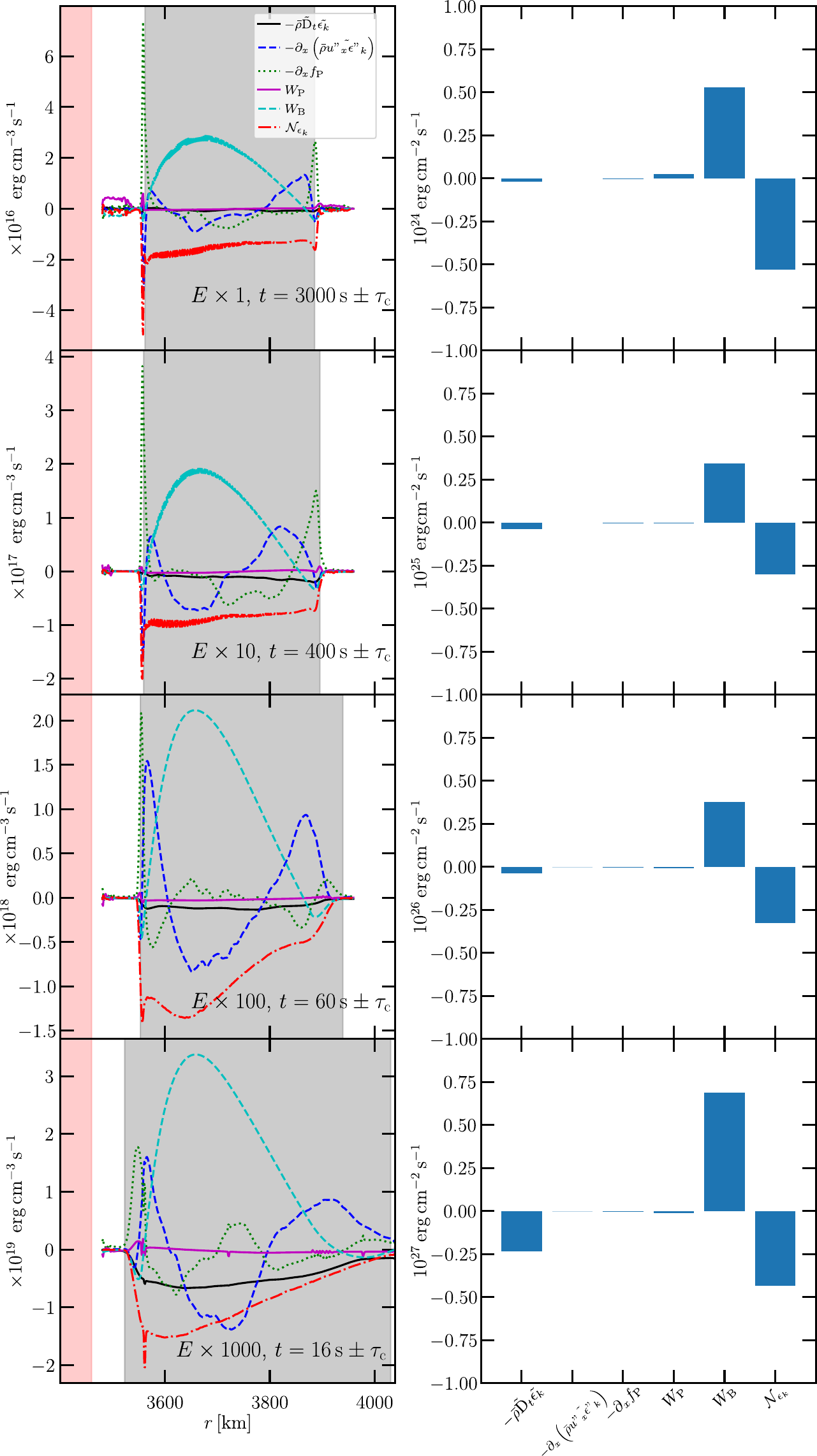}
\caption{Same as Fig.~\ref{fig_RANS_MKEE_reso}, but for various boosting factors instead of various resolutions. From top to bottom: nominal nuclear energy generation rate, energy generation rate boosted by a factor of 10, of 100, and of 1000. The time averaging is made over 2 turnover timescales centered on a time where the convective zone has about the same radial extent for each case (see text). The resolution of the simulations shown here is $512^3$.}
\label{fig_RANS_MKEE_boost}
\end{figure*}
\end{center}

In order to gain a better understanding of the key processes taking place in our simulations, we performed a mean field analysis, called RA-ILES (Reynolds-Averaged analysis of Implicit Large Eddy Simulations) hereinafter \citep[see][]{2007ApJ...667..448M,Viallet2013a,2017MNRAS.471..279C,Arnett2018b}. This allows us to disentangle the contributions and interplay between nuclear reactions and turbulence and to determine the dominant processes at play. It also provides insightful quantitative information. The basics of this analysis rely on a time- and space-averaging of all the quantities. The time averaging is done on a time-window $T$ which is long enough to be statistically meaningful, but short enough so that the main properties of the fluid do not change significantly. Usually, we perform this averaging over a time corresponding to two convective turnover timescales. The spatial averaging is performed over a given volume $V$ (in our simulation, it is usually the ``volume'', or rather surface of a horizontal slice of cells). One can therefore define the ``Reynolds average'' of a quantity $q$ as:
\begin{equation}
\bar{q}(x,t) = \frac{1}{T\Delta S}\int^{t+\frac{T}{2}}_{t-\frac{T}{2}}\int_{\Delta S} q(t',x,y,z) \text{d}S\text{d}t',
\end{equation}
$\text{d}S$ being an infinitesimal surface centered on the point $(x,y,z)$. The quantity $q$ can then be decomposed as:
\begin{equation}
q = \bar{q} + q'.
\end{equation}
Another useful type of averaging is a density-weighted average, also called ``Favre average'':
\begin{equation}
\Fav{q} = \frac{\Rey{\rho q}}{\Rey{\rho}}.
\end{equation}
Again, we can decompose any quantity $q$ as:
\begin{equation}
q = \Fav{q}+ q''.
\end{equation}
Note that most of the time, $q'\neq q''$.

With these notations, we can perform averages of the Euler equations governing the fluid we are simulating. Without providing details \citep[which can be found in\footnote{\footnotesize{For a general introduction to the RA-ILES method, see \citet{Chassaing2002a}.}}][]{Viallet2013a,Mocak2014a,Arnett2015a,Mocak2018a}, this leads to the following results for the kinetic energy equation:
\begin{equation}
\Rey{\rho}\,\Fav{\text{D}}_t\Fav{\epsilon}_\text{k} = -\nabla_x\Rey{\rho}\Fav{u_x''\epsilon_\text{k}} - \nabla_x f_\text{P} + W_P + W_\text{B}.\label{eq_kin}
\end{equation}
$\Rey{\rho}\,\Fav{\text{D}}_t$ is an operator similar to the Lagrangian derivative in the RA-ILES framework: $\overline{\rho}\Fav{\text{D}}_t \Fav{q} = \partial_t\left(\Rey{\rho}\Fav{q}\right) + \nabla_x \left(\Rey{\rho}\Fav{u}_x\Fav{q}\right)$, with $\nabla_x$ the $x$ component of the divergence operator\footnote{\footnotesize{In this work, $y$ and $z$ are both horizontal directions, while $x$ is directed outwards (opposite to the gravity field).}}. $f_\text{P} = \overline{P^\prime u^{\prime}_x}$ is the acoustic flux (i.e. the flux of pressure variations). $W_\text{P} = \overline{P'\partial_x u_x'}$ represents the turbulent pressure dilatation, and $W_\text{B} = -\overline{\rho}\overline{u_x''}\Fav{g}$
is the buoyancy work.
 
In this section, we go into the details of the RA-ILES decomposition of the turbulent kinetic energy (TKE) by discussing the importance and behaviour of each term in equation~(\ref{eq_kin}) for different resolutions ($128^3$ to $1024^3$ for the \texttt{Ex10} boosting factor, see Fig.~\ref{fig_RANS_MKEE_reso}) and boosting factors (nominal luminosity to $10^3$ times nominal luminosity at a resolution of $512^3$, see Fig.~\ref{fig_RANS_MKEE_boost}):

\textbf{Time Evolution} - $\Rey{\rho}\Fav{D_t}\Fav{\epsilon_k}$ represents the Lagrangian time derivatives of the kinetic energy. A negligible time derivative within the chosen time average implies that the convection is in a statistically steady state. This can seen to be true in all resolutions (Fig.~\ref{fig_RANS_MKEE_reso}) and most boosting factors. The only exception is the \texttt{Ex1000} model shown in Fig.~\ref{fig_RANS_MKEE_boost}. In this case, the high boosting factor leads to a rapid growth of the convective region, causing it to interact with the domain boundary before a steady state can be achieved. As such, we will not analyse the Reynolds averaged quantities of this model further.  

\textbf{Transport Terms} - The transport of kinetic energy throughout the convective region is defined by the TKE flux, $\nabla_x(\Rey{\rho} \Fav{u^{\prime\prime}_{x}\epsilon^{\prime\prime}_k)}$, and the acoustic flux, $\nabla_x f_P$, where $f_P = \overline{P^\prime u^{\prime}_x} $. We see that the acoustic flux often opposes the TKE flux within the convection zone, except at the convective boundaries, where the acoustic flux spikes where it launches gravity waves. The general behaviour observed in the \texttt{Ex10} case in Fig.~\ref{fig_RANS_MKEE_reso} can also be seen when the boosting factor varies (Fig.~\ref{fig_RANS_MKEE_boost}). The general pattern is preserved, but its amplitude is increased for increasing boosting factors, since the higher velocity of the fluid makes the transport more efficient (note the change in the scale on the y axis by a factor of 1000 from the top to the bottom panel).

\textbf{Source Terms} - Turbulence in the convective region is driven by the work due to turbulent pressure fluctuations, $\mathbf{W}_P$ and the buoyancy work due to density fluctuations, $\mathbf{W}_B$. As in previous work by \cite{Viallet2013a} and \cite{2017MNRAS.471..279C}, $\mathbf{W}_P$ appears to be negligible in shell convection in deep interiors. $\mathbf{W}_B \, > \,0$ is generally seen in the convective zone as expected, since it is the main driving term. Regions near the convective boundary have $\mathbf{W}_B \, < \,0$, meaning that the flow decelerates in these regions. These regions are usually referred to as ``overshooting'' regions in 1D stellar modelling but we prefer the more general term of ``convective boundary mixing'' regions. We notice a minor systematic decrease in the integrated buoyancy work as resolution is increased. We suspect this effect is caused by the lower resolution models having a larger convective boundary mixing extent. This leads to a higher rate of entrainment of new material to burn into the convective Ne shell, which in turn, slightly increases the rate of energy generation and hence the buoyancy work done. The difference, however, is quite minor even over the relatively long timescales that we simulate. Looking at the different boosting factors (Fig.~\ref{fig_RANS_MKEE_boost}), we see that $\mathbf{W}_B$ scales linearly with them: the work done by buoyancy force is about 1000 times higher in the \texttt{Ex1000} simulation compared to the nominal case \texttt{Ex1}. This is due to the fact that nuclear processes drive convection by heating up the plasma at the bottom of the shell.

\textbf{Dissipation} - Due to the finite size of our grid, our code is not able to perfectly reproduce the behaviour of the fluid at spatial sizes which are smaller than the grid size and our simulations do not include explicit viscosity (ILES). The numerical dissipation (taking place at the grid scale) is nevertheless tracked by the $\mathcal{N}_{\epsilon_k}$ term, which is the difference between the left-hand side and right-hand side terms of Eq.~\ref{eq_kin}. Similar to the results found in \cite{2017MNRAS.471..279C} for a convective carbon burning shell, we find that the work due to buoyancy, $\mathbf{W}_B$, is closely balanced by numerical dissipation $\mathcal{N}_{\epsilon_k}$ as expected from Kolmogorov's turbulence theory as numerical dissipation in ILES replaces the dissipation due to physical viscosity. One key feature that depends on resolution is the numerical dissipation peak at the lower boundary. This confirms that high resolution is needed to fully resolve the lower boundary of convective region. Another feature that affects low resolution simulation is the leap-frog instability that creates zig-zags in the $\mathbf{W}_B$ and thus also the $\mathcal{N}_{\epsilon_k}$ profiles. This analysis shows that a resolution of 256 or higher is best to resolve the key processes taking place in the bulk of convective regions and not-too-steep boundaries like the top boundaries in this study. A resolution of 1024 or higher is needed to perfectly resolve sharp boundaries like the lower ones but measurements of entrainment do not necessarily require the lower boundary to be perfectly resolved \citep{Rizzuti2022b,Rizzuti2023b}.

\begin{figure}
\centering
\footnotesize
\includegraphics[width=0.5\textwidth]{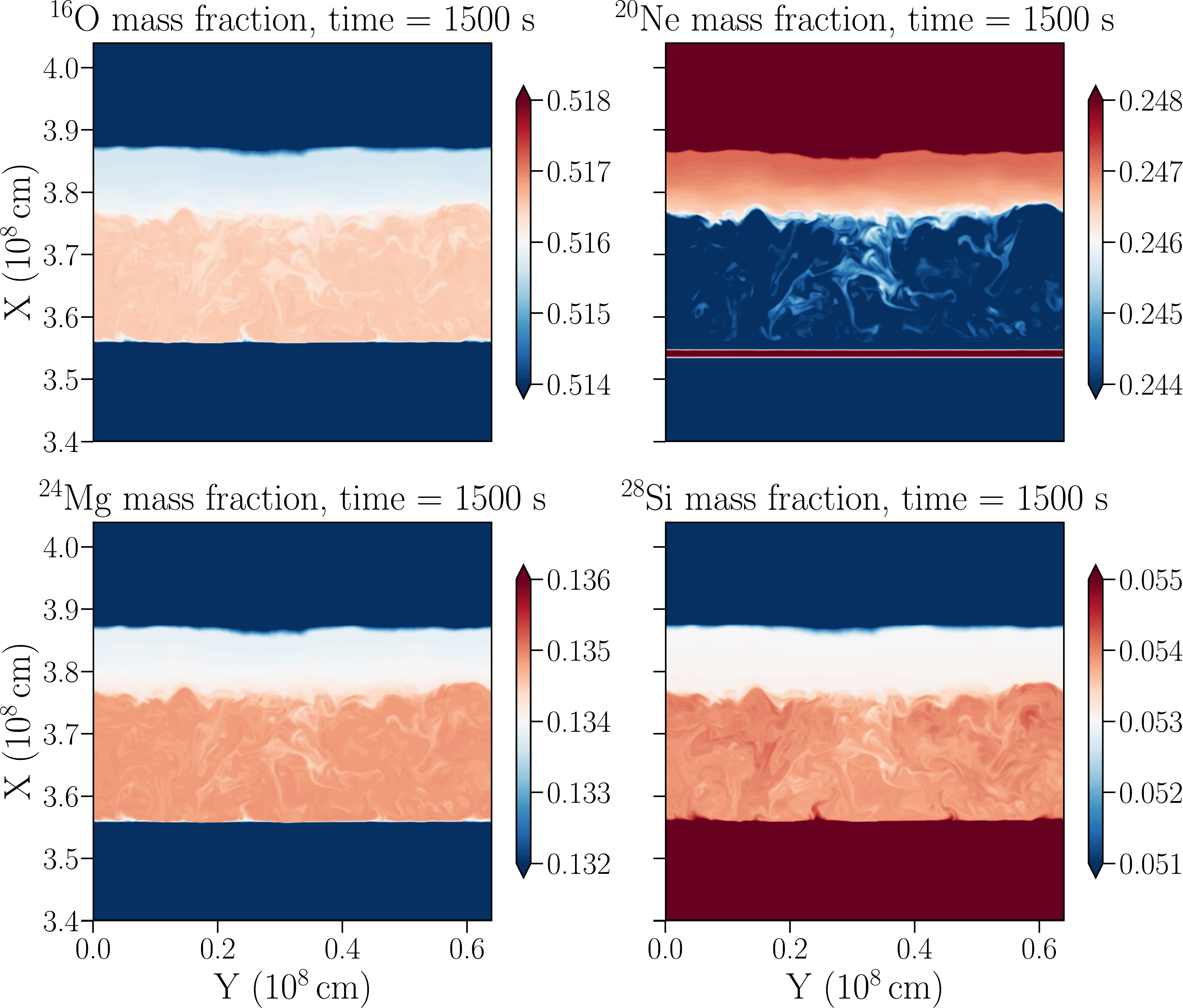}
\caption{Vertical cross sections of the mass fraction (values in colour scale) of the different isotopic nuclei $^{16}$O, $^{20}$Ne, $^{24}$Mg, $^{28}$Si, taken at 1500 seconds in the \texttt{Ex1} model. Reference value of the colourbars (white) is the average mass fraction of each nuclides in the convective zone at the beginning of the simulation. For this reason, the overabundance of $^{16}$O, $^{24}$Mg, $^{28}$Si (toward the red) and the underabundance of $^{20}$Ne (toward the blue) inside the convective zone reflect the nuclear reactions which are burning neon to produce oxygen and magnesium, and burning magnesium to produce silicon.}\label{fig_mass_cross_section}
\end{figure}

\begin{figure*}
\footnotesize
\includegraphics[width=0.48\textwidth]{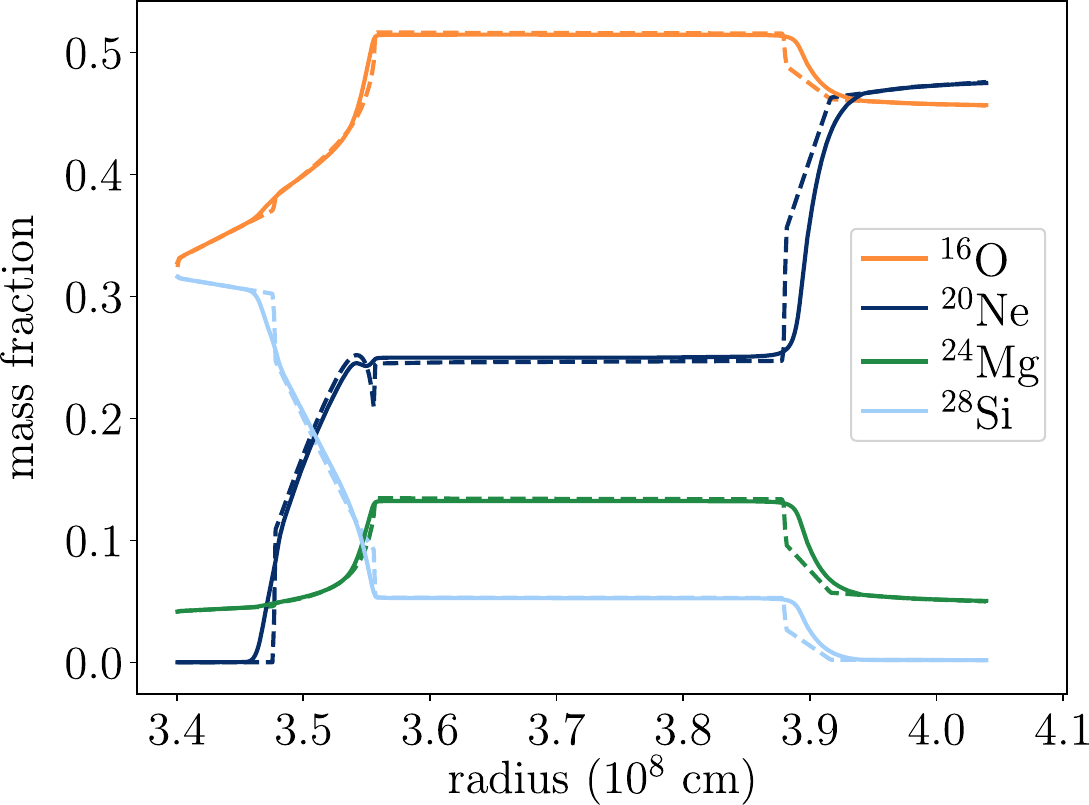}
\includegraphics[width=0.48\textwidth]{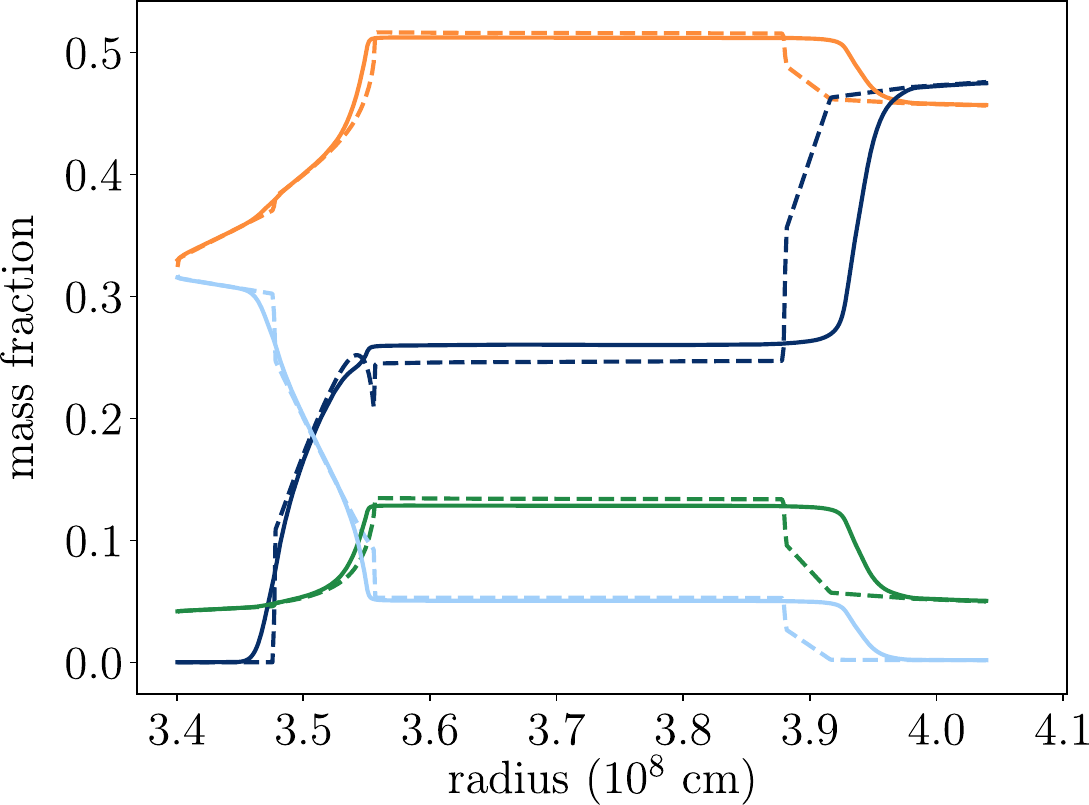}\\
\includegraphics[width=0.48\textwidth]{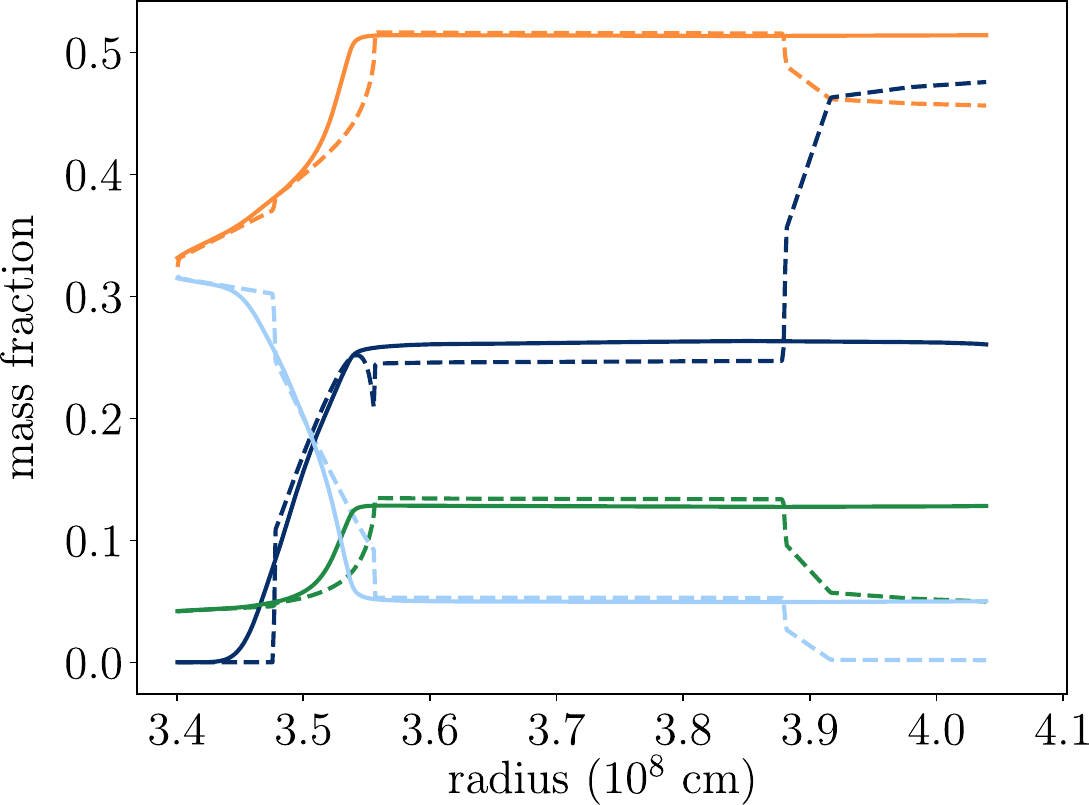}
\includegraphics[width=0.48\textwidth]{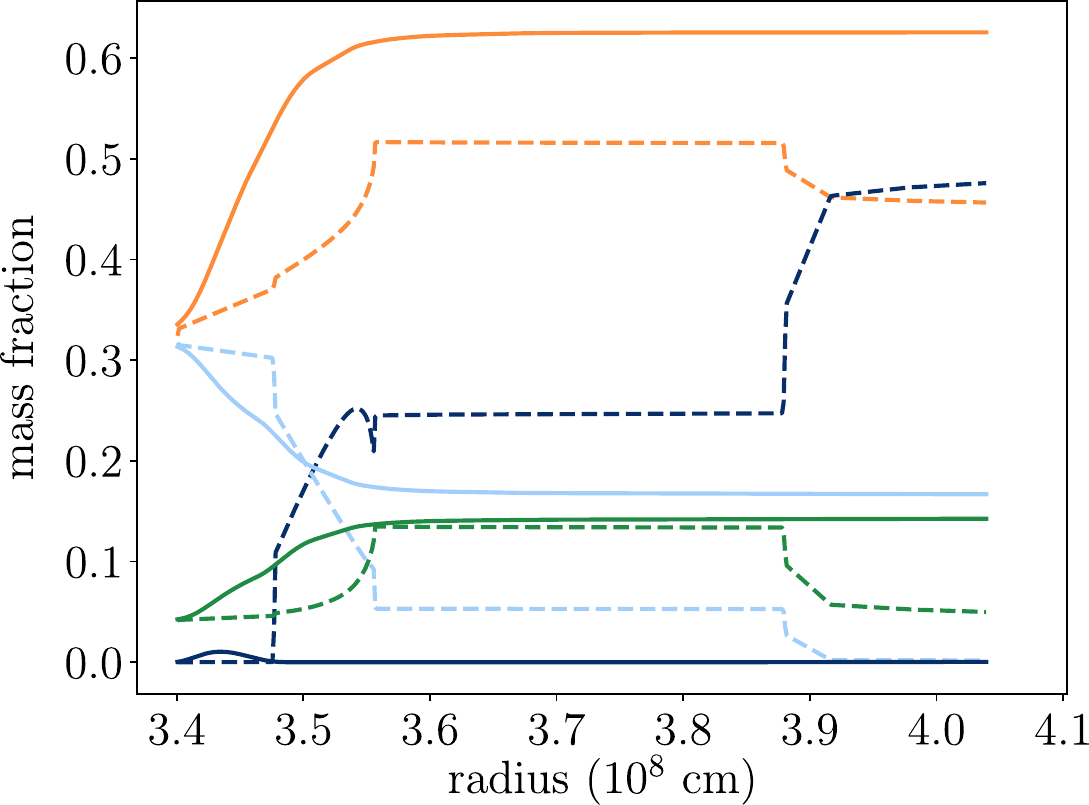}
\caption{Profiles of the abundances of the four chemical elements followed in our set of simulations at the beginning (dashed lines) and at the end (solid lines) of the simulated time, and for \texttt{Ex1} (top-left), \texttt{Ex10} (top-right), \texttt{Ex100} (bottom-left), and \texttt{Ex1000} run (bottom-right).}\label{fig_14}
\end{figure*}

\section{Interplay between turbulence and nuclear processes}\label{nuclear}
Our simulations follow explicitly the evolution of the four key nuclides for neon burning: $^{16}$O, $^{20}$Ne, $^{24}$Mg, $^{28}$Si, which enables us to study the interplay between turbulence and nuclear processes in great details. These nuclides are linked by a small tailored nuclear reaction network including the following reactions: $^{20}$Ne$(\gamma,\alpha)^{16}$O, $^{16}$O$(\alpha,\gamma)^{20}$Ne, $^{20}$Ne$(\alpha,\gamma)^{24}$Mg, $^{24}$Mg$(\alpha,\gamma)^{28}$Si ($\alpha$ particles are considered to be at nuclear equilibrium, which is a reasonable assumption for the neon shell studied here).

 We can see the effects of the above-mentioned neon-burning reactions in the vertical cross-section snapshots presented in Fig.~\ref{fig_mass_cross_section}, where we show in colour scale the mass fraction of these four species after 1500 seconds from the start of the \texttt{Ex1} model (note that the simulation has not reached the quasi-steady-state by that time and that the turbulent region has not reached the upper boundary from the initial 1D model). In particular, the reference colour (white) indicates the average mass fraction inside the neon shell at the beginning of the simulation. The colour represents the same range of deviations from the initial composition in the convective region in all panels. The results fully reflect what is expected for neon burning: neon is consumed inside the convective region, while oxygen, magnesium and silicon are produced via the reactions listed above. These abundance plots also show that the species are not always completely homogeneously mixed and that they can be used as tracers of the turbulent motion of the fluid. Particularly interesting is the $^{20}$Ne case: the bottom part of the convective region shows an under-abundance with respect to the initial one. This is a sign of the nuclear burning taking place at the bottom boundary. On the contrary, the top part of the convective region shows an overabundance of neon. This is due to entrainement, which brings neon from the stable region above the convective zone (where abundance is higher) into the convective zone, where it is slowly mixed. Note that the strong overabundance in Ne between $3.5$ and $3.6\cdot 10^8\,\text{cm}$ comes from the initial 1D structure (see Fig.~\ref{fig_14}, dashed line), which has not been erased by turbulent mixing at the time when the snapshot is taken here, but it is erased soon afterwards.

\subsection{Evolution of the chemical composition}

An overview of the time evolution of the chemical composition in our simulations is presented in Fig.~\ref{fig_14}, which shows the radial profiles of the four chemical elements followed in our simulations (i.e. the horizontal average of the abundance) at the start of the simulation (dashed lines, the same in each case), and at the end (solid lines), for the four boosting factors. The convective region corresponds to the plateaus in the middle of the computational domain. A few general observations can already be made from this figure. The effects of entrainment are clearly visible, particularly at the top boundary: the boundary of the convective zone has moved outwards, and the steep initial profiles have been smoothed by convective boundary mixing. Interestingly, in the current set of simulations, entrainment is more efficient at bringing $^{20}$Ne inside the convective zone than nuclear burning at destroying it, leading to a slight increase in $^{20}$Ne abundance (except for the \texttt{Ex1000} case). Qualitatively, the \texttt{Ex1} and \texttt{Ex10} cases are very similar, except that, as expected, entrainment is more efficient with the higher boosting factor. This is exacerbated for the \texttt{Ex100} and \texttt{Ex1000} cases where the convective boundary has moved up to the top of the computing domain over the course of the simulation.

\subsection{RA-ILES analysis of the composition transport equation}
As was done for the turbulent kinetic energy in the previous section, our RA-ILES can also be applied to the chemical composition transport equation, resulting in the following equation:
\begin{equation}
\Rey{\rho}\Fav{\text{D}}_t\Fav{X}_i = -\nabla_x f_i + \Rey{\rho}\Fav{\dot{X}}^{\text{nuc}}_i.\label{eq_chem}
\end{equation}
$X_i$ is the mass fraction of the chemical element $i$, $f_i = \Rey{\rho}\Fav{X_i''u_x''}$ is the turbulent flux of the element $i$, and $\dot{X}_i$ is the rate of creation/annihilation of the element $i$ due to nuclear reactions. The right-hand side of this equation includes the two sources of the modification of the mass fraction at a given location inside the computing domain: turbulent motion transport and nuclear processes.

\begin{figure}
\includegraphics[width=.5\textwidth]{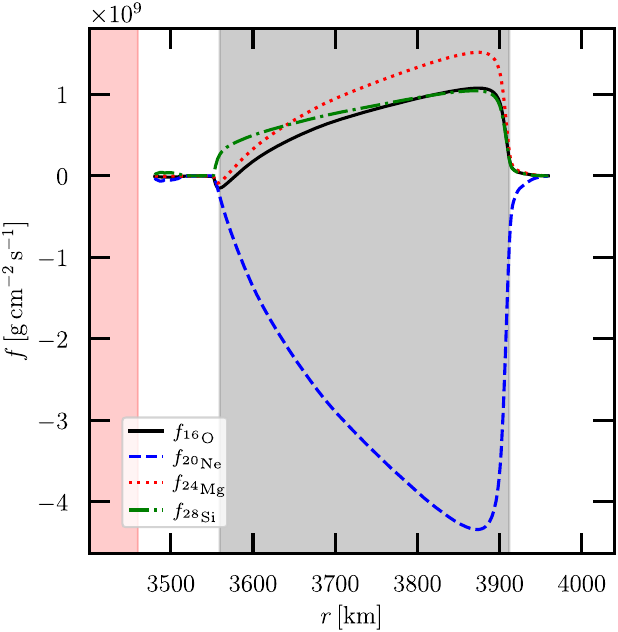}
\caption{Turbulent composition fluxes for the nuclides explicitly followed in our work. These fluxes are taken from the $512^3$ resolution simulation with a boosting factor of 10. The time averaging is done as in Fig.~\ref{fig_RANS_MKEE_reso}, and the coloured areas have the same meaning. The flux for each element is defined as 
$f_i = \Rey{\rho}\Fav{X_i''u_x''}$
\citep[cf.][]{Mocak2018a}.}
\label{fig_RANS_ChemFlux}
\end{figure}

Before discussing the composition transport equation, it is useful to understand the turbulent composition flux, $f_i$, plotted in Fig.~\ref{fig_RANS_ChemFlux}, for $^{16}\text{O}, ^{20}\text{Ne}, ^{24}\text{Mg}$ and $^{28}\text{Si}$. These fluxes represent the rate at which each species is transported. Negative values indicate that the fluxes are oriented inward. We see that large quantities of neon is being entrained from the upper boundary, and transported towards the inner parts of the convective burning zone, where the neon is burnt. The other species, on the other hand, are being transported outward (to the right of the convective zone). The behaviour at the inner boundary is more complex, due to the interplay between entrainment through the boundary, and the burning occurring slightly above. $^{16}\text{O}$ and $^{24}\text{Mg}$ are more abundant inside the convective zone than outside (see Fig.~\ref{fig_14}). There is thus a slightly negative flux near the bottom boundary for these elements, showing that they are partly transported outside the convective region. On the other hand, both elements are produced through the burning of $^{20}\text{Ne}$. Convective motions transport the freshly produced elements through the convective zone, as indicated by the positive flux inside the bulk of the convective region. The case of $^{28}\text{Si}$ is different still: there is more of it below the convective zone than inside. It is thus transported through the bottom boundary by entrainment into the convective region, hence the positive flux at the boundary. Furthermore, this element is also synthesised by neon burning (by a double $\alpha$-capture), and redistributed inside the whole convective region from bottom to top, hence the positive flux for $^{28}\text{Si}$ everywhere.

\begin{figure*}
\includegraphics[width=\textwidth]{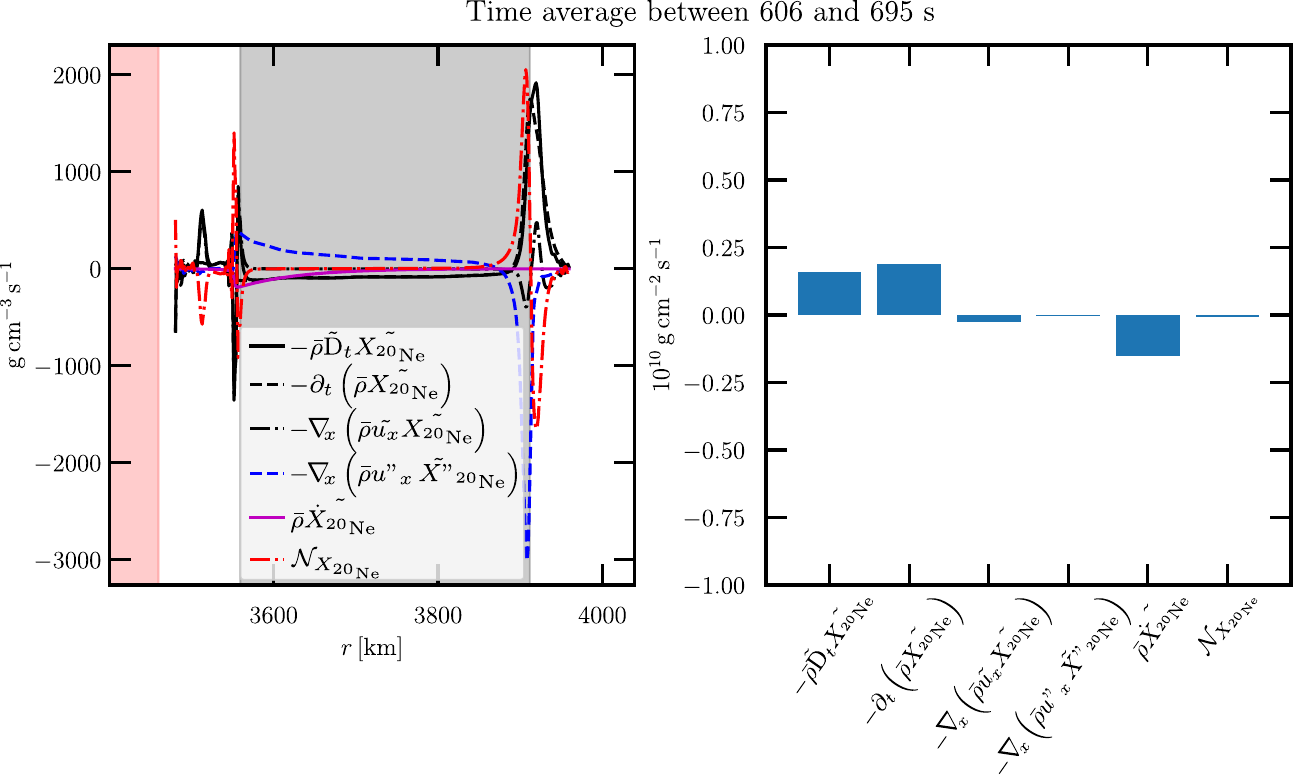}
\caption{\textit{left panel:} RA-ILES decomposition of the mean $^{20}$Ne abundance equation. The meaning of each curve is specified in the legend. The coloured regions have the same meaning as in Fig.~\ref{fig_RANS_MKEE_reso}, and the time average is performed over the same time window. The results come from the $512^3$ resolution simulation, with a boosting factor of 10. \textit{Right panel:} Radial integration of each term shown on the left side of the figure.} 
\label{fig_RANS_Ne20}
\end{figure*}

\begin{figure}
\includegraphics[width=.5\textwidth]{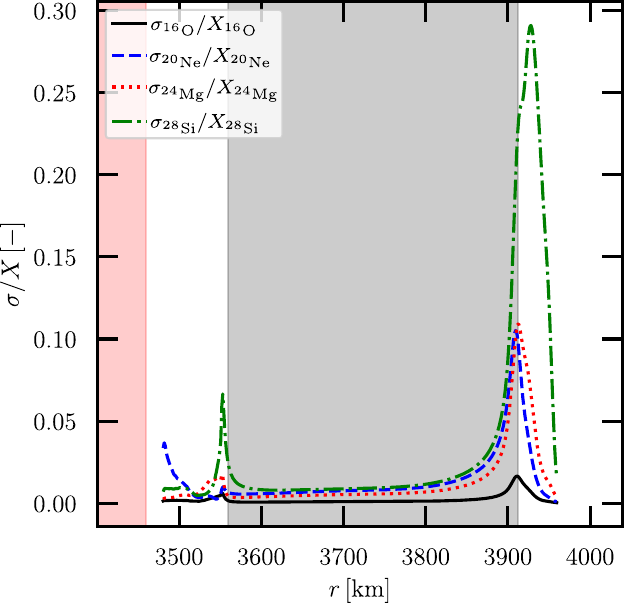}
\caption{Standard deviation of a chemical element abundance, normalised by its abundance. The coloured regions have the same meaning as in Fig.~\ref{fig_RANS_MKEE_reso}, and the time average is performed over the same time window. The results come from the $512^3$ resolution simulation, with a boosting factor of 10.}
\label{fig_RANS_Ne20Variance}
\end{figure}

The RA-ILES composition transport equation profiles for neon are shown in Fig.~\ref{fig_RANS_Ne20}. Each term of eq.~\ref{eq_chem} is shown by a different curve in the plot and is described below:

\textbf{Time Evolution} - $\Bar{\rho}\Fav{D_t}\Fav{X}_{i}$ and $\partial_t(\Bar{\rho}\Fav{X}_{i})$ represents the Lagrangian and Eulerian time derivatives of the composition. While the convective region is in a steady state, we see that significant mixing occurs at the convective boundaries, particularly at the upper boundary, which changes the size of the convection zone over time.  

\textbf{Transport Terms} - The mean composition flux term is given by $\nabla_x(\Bar{\rho}\Fav{u_x}\Fav{X}_{i})$, which we see is non-negligible at the upper convective boundary, implying that there is a change in the composition profile due to the overall growth of the convective zone. This is due to a mean flux of Ne into the convective zone from the upper boundary, as seen in Fig.~\ref{fig_RANS_ChemFlux}.  The divergence of the turbulent flux, $\nabla_x(\Rey{\rho}\Fav{X_i''u_x''})$, shows how material mixed into the convective zone is transported by turbulent velocity variations. This term is positive where the abundance is decreased by the element flux, and negative where there is an accumulation of a chemical element due to transport. We see in Fig.~\ref{fig_RANS_Ne20} that the turbulent fluxes accumulate the neon abundance near the top boundary. The time rate of change of neon balances this term. At the bottom boundary, the destruction of neon by the nuclear burning is compensated by neon being transported here.

\textbf{Nuclear burning} - The rate at which compositions change due to nuclear burning, $\Bar{\rho}\Fav{\Dot{X}}_{i}$, shows how the rate of neon burning increases with depth, and hence density and temperature. As it also depends on the abundance of the fuel, this term peaks at the bottom of the convective zone. 

\textbf{Numerical residual} - $\mathcal{N}_{X_i}$ - This term highlights the loss of information of our code at the grid level. It is small throughout most of our simulated domain, but we see that it is important near the edge of the convective zone, where strong turbulence develops. Increasing the resolution of the simulation makes this term smaller \citep[][and Fig. \ref{fig_RANS_MKEE_reso}]{Arnett2018b}.

Integrated over the whole computational domain (\textit{right panel}), we see that the global change of the $^{20}\text{Ne}$ abundance is, as expected, mostly due to the nuclear reactions, which are responsible for the decrease in the abundance of this chemical element.

\begin{figure*}
\includegraphics[width=\textwidth]{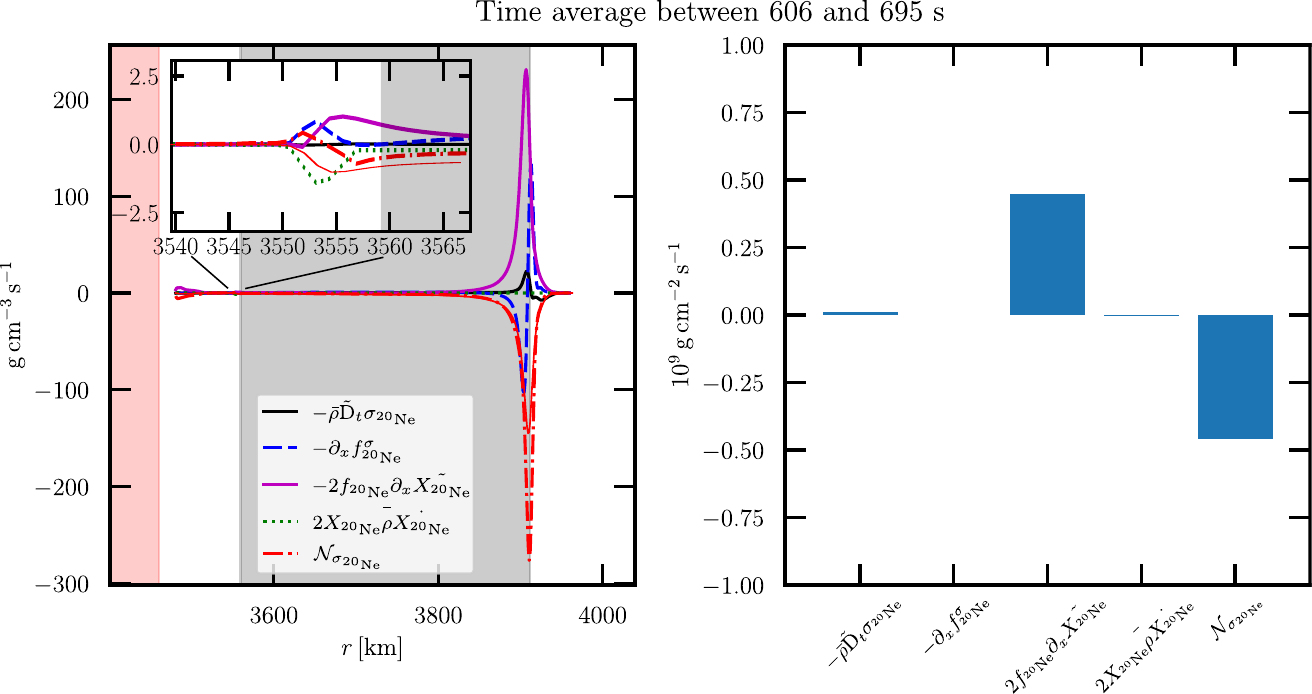}
\caption{\textit{left panel:} RA-ILES decomposition of the mean $^{20}$Ne abundance variance equation. The meaning of each curve is specified in the legend. The coloured regions have the same meaning as in Fig.~\ref{fig_RANS_MKEE_reso}, and the time average is performed over the same time window. The results come from the $512^3$ resolution simulation, with a boosting factor of 10. The thin red line shows a modelling of the residuals, done as in \citet{Mocak2018a}. \textit{Right panel:} Radial integration of each term shown on the left side of the figure.}
\label{fig_RANS_Ne20Var}
\end{figure*}

\begin{figure}
\includegraphics[width=.5\textwidth]{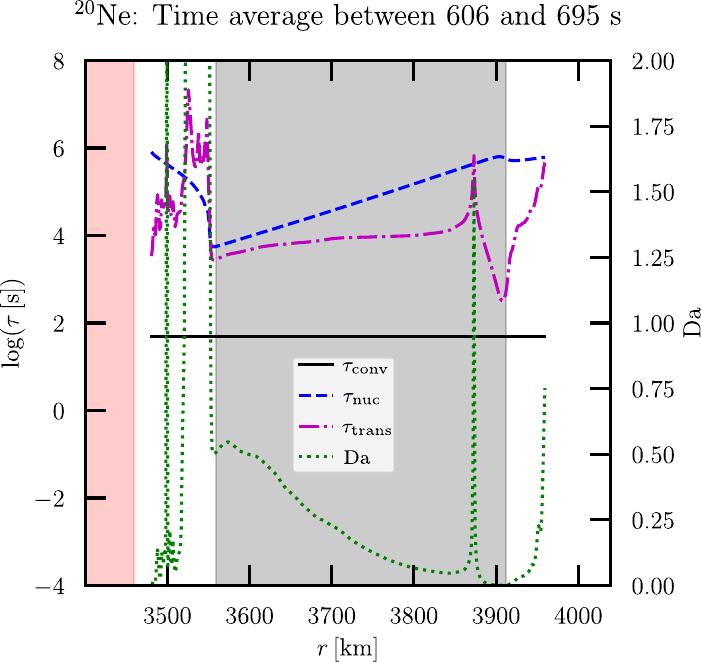}
\caption{Characteristic times for the $^{20}$Ne. The convective turnover timescale is shown by the solid black line. The nuclear timescale is shown in dashed blue. The transport timescale is shown in dot-dashed purple. The Damkh\"oler number is shown in dotted green (value to be read on the right axis). The coloured regions have the same meaning as in Fig.~\ref{fig_RANS_MKEE_reso}, and the time average is performed over the same time window. The results come from the $512^3$ resolution simulation, with a boosting factor of 10.}
\label{fig_RANS_Ne20Tau}
\end{figure}

\subsection{RA-ILES analysis of the composition variance}
The variance, $\sigma$, of the composition traces the variation of the abundance of an element with respect to its average value at a given radius inside the computational domain. In our RA-ILES framework, it is defined as $\sigma_X = \Fav{X''X''}$. In Fig.~\ref{fig_RANS_Ne20Variance} we show the relative standard deviation of the composition, i.e. $\frac{\sqrt{\sigma_X}}{\Rey{X}}$. In the bulk of the convective region, the deviation remains small (around 1\%), justifying the general 1D modelling approach used in stellar evolution. We note two locations however where this deviation becomes larger for all the chemical elements: both boundaries of the convective zone. The convective boundaries are not flat or regular, but distorted and turbulent, making important variations with respect to the average of the abundance of a given element. This behaviour is much more important at the top boundary, where the relative deviation can reach about 30\% for silicon.

We can investigate the origins of the variance using the RA-ILES chemical variance equation:
\begin{equation}
\Rey{\rho}\Fav{\text{D}}_t\sigma_i = -\nabla_x f_i^\sigma - 2f_i\partial_x\Fav{X}_i + 2\Rey{X_i''\rho\dot{X}_i^\text{nuc}}.\label{eq_var}
\end{equation}
$\sigma_i = \Fav{X_i'' X_i''}$ is the composition variance of the element $i$, $f_i^\sigma = \Rey{\rho X_i'' X_i''u_x''}$ is the turbulent flux of the composition variance. $2f_i\partial_x\Fav{X}_i$ is a source term linked to the flux of the chemical element $i$, and $2\Rey{X_i''\rho\dot{X}_i^\text{nuc}}$ is a source term linked to nuclear reactions.

We discuss below the terms of Eq.~(\ref{eq_var}), which are presented in Fig.~\ref{fig_RANS_Ne20Var}:

\textbf{Time Evolution} - $\Bar{\rho}\Fav{D_t}\Fav{\sigma}_{i}$ (solid black line) and $\partial_t(\Bar{\rho}\Fav{\sigma}_{i})$ (dashed black line) represents the Lagrangian and Eulerian time derivatives of the composition variance. The two curves are almost identical throughout the domain, and are close to $0$ everywhere, except near the top convective boundary. At this location, entrainment of $^{20}\text{Ne}$ inside the convective region is a source of composition variance.

\textbf{Transport Terms} - $\nabla_x({f_i}^\sigma)$, is the turbulent composition variance flux, and is responsible for the redistribution of the composition variance inside the domain. As shown in Fig.~\ref{fig_RANS_Ne20Var} (see the blue curve), the variance produced near the top convective boundary is transported in both directions, outwards inside the stable radiative layer and inside the convective region. It progressively builds the variance peak visible in Fig.~\ref{fig_RANS_Ne20Variance}.

\textbf{Turbulent Production} - $2 f_i \partial_x(\Fav{X_i})$. This term is shown in magenta in Fig.~\ref{fig_RANS_Ne20Var}, and is the dominant term affecting the creation of composition variance. As this term is effective only in regions where a composition gradient and a turbulent composition flux exist at the same time, it is mostly present near the top convective boundary (and to a much smaller extent near the bottom one). Looking at the integrated values (right panel), one clearly sees that composition variance inside the domain is mostly built by this interaction between the composition flux and the composition gradient.

\textbf{Nuclear Burning} - The rate at which composition variance changes due to nuclear burning is traced by $2\Rey{X_i''\rho\Dot{X}_{i}}$. In our case, it only occurs at a quite low level near the bottom boundary of the convective zone (see the green curve in the zoom-in on this region), where nuclear burning takes place. This term is negative, indicating that nuclear burning decreases the composition variance.

\textbf{Numerical residual} - $\mathcal{N}_{\sigma_i}$ represents the residual in our simulation, i.\,e. the sum of all the other terms (with the same sign as shown in the legend of Fig.~\ref{fig_RANS_Ne20Var}). As the resolution of our simulation is finite and is therefore limited in spatial resolution, it does not reproduce the exact behaviour of the flow at very small scales, where turbulence is dissipated. The residual thus represents the dissipation at the grid level in our simulations. Despite its numerical origin, the behaviour of this dissipation term is  appropriate: we added on the left panel of Fig.~\ref{fig_RANS_Ne20Var} a theoretical modelling of this dissipation (thin solid red line), which fits well the residual obtained in our simulations. We adopted the same model as in \citet{Mocak2018a}, which assumes that the dissipation time-scale is the same as the Kolmogorov damping time-scale.

Here also, the radial integration over the whole domain of the previous terms provide insight into the main mechanisms leading to the change of the variance (here the $^{20}\text{Ne}$ one). We see that the turbulent transport is building the chemical variance inside the domain, while the dissipation at the grid level compensate for this net production.

\subsection{Characteristic time scales}
Another way to study the interplay and relative importance of turbulence and nuclear processes is by using characteristic timescales. The following characteristic time scales are relevant in this context:

\begin{enumerate}
    \item  The convective turnover time scale $\tau_\text{conv}$, which is defined as the typical time needed for a fluid elements to cross twice the convective zone (a ``convective loop''). We thus have
\begin{equation}
\tau_\text{conv} = \frac{2d_\text{conv}}{v_\text{rms}},
\end{equation}
where $d_\text{conv}$ is the radial extend of the convective zone and $v_\text{rms}$ is the average root-mean-square velocity inside the convective zone.
\item  The local nuclear burning time scale $\tau_\text{nuc}$, which is the typical time scale for nuclear burning to take place at a given location inside the star:
\begin{equation}
\tau_{\text{nuc},i} = \frac{X_i}{\dot{X_i}},
\end{equation}
where as above $X_i$ is the mass fraction of a given chemical element and $\dot{X}_i$ is the rate of creation/annihilation of the element $i$ due to nuclear reactions.

\item The local transport time scale $\tau_\text{trans}$, which is the characteristic time for transport to remove/bring a chemical element at a given place:
\begin{equation}
\tau_{\text{trans},i} = \frac{\bar{\rho}\Fav{X_i}}{\partial_x f_i},
\end{equation}
where $f_i$ is the flux of the chemical species.
\end{enumerate}

Based on these characteristic time scales, one can define the so-called Damk\"ohler number as the ratio between turbulent transport and nuclear time scales for a given chemical element:
\begin{equation}
\text{Da}_i = \frac{\tau_{\text{trans},i}}{\tau_{\text{nucl},i}},
\end{equation}
with the timescales defined above. 
The Damk\"ohler number discriminates between two different regimes inside the convective zone: when $\text{Da}$ is smaller than one, transport timescale is smaller than the nuclear burning timescale. Chemical species are thus well mixed inside the convective region and inhomogeneities (illustrated by the variance) remain small inside the convective zone (except near the boundaries, as explained above). This is how the convective zones look like in 1D stellar evolution. However, it is possible in some cases that the Damk\"ohler number becomes close or even larger than one. In this case, nuclear burning is faster than the transport, and the mixing inside the convective zone is less efficient. This is the ``convective-reactive regime'' discussed in \citet{Herwig2011a} \citep[see also][]{Mocak2018a}. 

The characteristic time scales and the Damk\"ohler number are shown in Fig.~\ref{fig_RANS_Ne20Tau} for $^{20}\text{Ne}$ for our \texttt{Ex10} simulation. The fastest of the time scales is the convective time scale (solid black line), which is slightly less than $100\,\text{s}$. 
The transport timescale (magenta dot-dashed line) is roughly constant over a significant fraction of the convective zone (about $10^4\,\text{s}$), with a peak at the location where the gradient of flux vanishes (see Fig.~\ref{fig_RANS_ChemFlux}), and has a smaller value near the top boundary (less than $10^3\,\text{s}$).
Finally, the nuclear burning timescale (blue dashed line) is shortest where nuclear burning is strongest, at the bottom of the convective zone (about $10^4\,\text{s}$, and increases outwards, to reach $10^6\,\text{s}$ near the top boundary). The Damk\"ohler number (green dotted line, scale on the right) is thus significantly smaller than one everywhere inside the convective region. This means that the mixing of $^{20}\text{Ne}$ is faster than nuclear burning, and that the chemical composition of the convective region (at least concerning $^{20}\text{Ne}$, but it is true also for the other chemical species followed in this work) is well homogenised. The same is true for the nominal luminosity case, which means that 1D stellar models are sufficient to reproduce standard neon burning (the same is true for the \texttt{Ex10} and \texttt{Ex100}).

\begin{figure*}
\includegraphics[width=.45\textwidth]{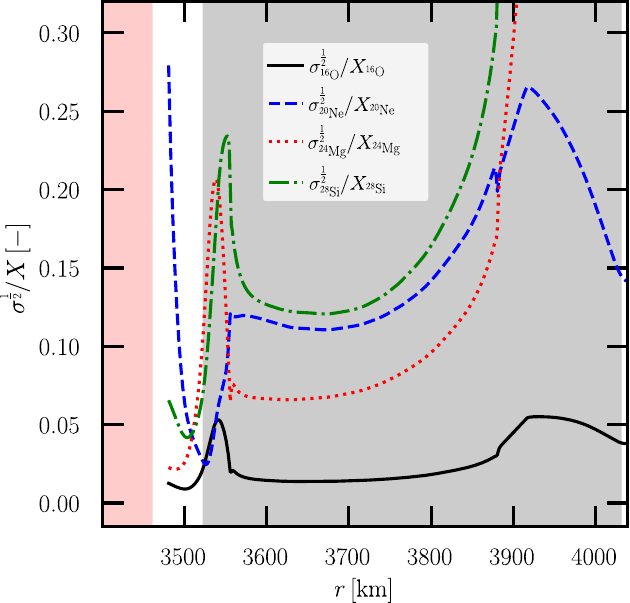}\hfill\includegraphics[width=.5\textwidth]{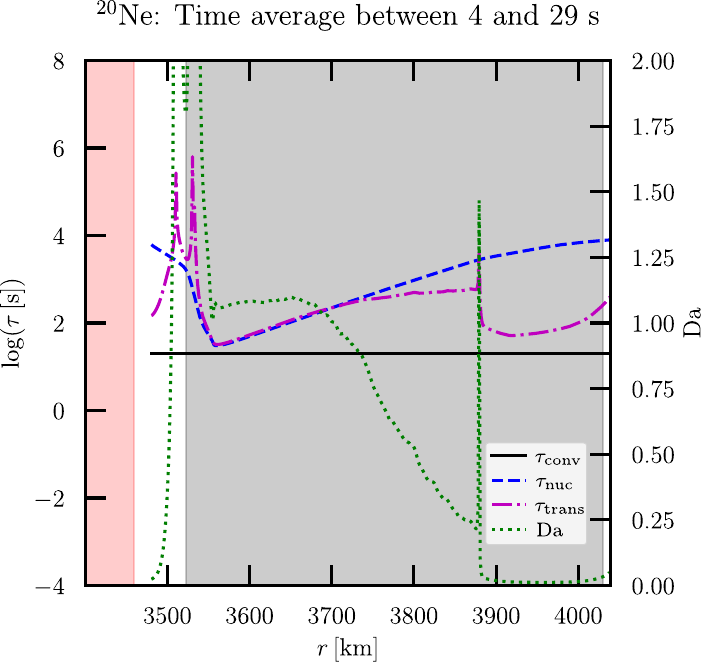}
\caption{\textit{Left panel:} relative variance for the 4 elements followed in this study, in the case of the \texttt{Ex1000} simulation. \textit{Right panel:} Characteristic timescales and Damköhler number for the $^{20}\text{Ne}$, in the case of the \texttt{Ex1000} simulation. For both panel, the quantities are averaged between $4$ and $29\,\text{s}$.}
\label{fig_Ne20Var_and_Tau}
\end{figure*}

The \texttt{Ex1000} simulation, on the other hand, behaves very differently. This extreme case is illustrated in Fig.~\ref{fig_Ne20Var_and_Tau} for $^{20}\text{Ne}$ ({\it right panel}). Variances ({\it left panel}) are about 10 times larger than in the \texttt{Ex10} case shown in Fig.~\ref{fig_RANS_Ne20Variance} in the whole convective region. In the inner part of the convective zone, nuclear and transport timescales are about the same, with Damk\"ohler number equal or larger than one. Looking at Fig.~\ref{fig_14}, we see that the composition profiles at the end of the simulation (solid lines) for the \texttt{Ex1000} case are not perfectly flat, but show a gradient in the inner part of the convective zone, as expected from the high Da number. In such cases, nuclear burning and convective mixing can still be treated correctly in 1D models if for example these two processes are solved in a coupled manner as done e.\,g. in the MESA code \citep{2011ApJS..192....3P}. Non-negligible variances, however, cannot be reproduced in 1D models and such cases need 3D simulations to simulate all the details. While the \texttt{Ex1000} simulation is not representative of neon burning inside massive stars, there are several cases where we expect Da$\sim 1$. In addition to the two cases mentioned above \citep{Herwig2011a,Mocak2018a}, other ``convective-reactive'' environments are expected in various H-He shell interaction events \citep{Hirschi2007, Clarkson2021} and shell merger situations already found in 1D models \citep{Rauscher2002, Tur2007, Ritter2018, Cote2020}, and in the 3D pre-supernova simulation of \citet{Yadav2020}.

\section{Discussion and conclusions}\label{Conclu}

This paper focuses on the 3D hydrodynamics study of a neon-burning shell during the advanced stages of the life of a massive star. We performed a large set of simulations with the PROMPI code at different resolutions (from $128^3$ to $1024^3$ cells), and with different boosting factors for the nuclear energy generation rate (from nominal to 1000 times boosted energy generation rate). The simulations were followed for a long enough time to ensure the flow has significant statistical properties (generally more than 5 turnover timescales).

Our simulation at a nominal luminosity allows us to draw conclusions about the flow without any extrapolations. In the initial phases, we find that the convective velocities found in 3D are $\approx 2$ times larger than those expected from MLT approximations. Due to CBM at the convective boundaries, over the course of the simulation (about 7 convective turnovers),
 the convective shell grows noticeably, and the extra fuel entrained leads to an increase in the convective velocity. 

A comparison of different resolutions shows that even at the lowest resolution ($128^3$), the large-scale convective flows are reasonably resolved. They attain comparable values of kinetic energy and similar turbulent cascade in the velocity spectra at large wavelengths to their high-resolution counterparts. Using our RA-ILES analysis framework, we also performed a thorough comparison of the mean-field equations at the different available resolutions. As expected from previous works \citep{Viallet2013a, 2017MNRAS.471..279C}, an analysis of the mean turbulent kinetic energy equation confirms that the statistical properties of the flow are already well captured by the simulation at a relatively low resolution. 
Higher resolutions, however,  are necessary to resolve the convective boundaries and the corresponding mixing across them.

We analyse the mean field composition variance equation, which allows us to determine the degree of deviation of the 3D simulations from 1D averages. We find that the deviation is small (around $1\%$) in most of the convection zone. At the convective boundary, matter is entrained from the stable region inside the convective region. The boundaries are less homogeneous than the rest of the convective zone, and the variance (in terms of mass fraction) can become as high as about 30\% (for silicon). Our mean-field analyses clearly show a net production of variance near the boundaries. This result justifies the general 1D stellar modelling approach for convection but reiterates the need for stellar modellers to use a CBM prescription at convective boundaries. 

The various boosting factors we used allowed us to determine how different quantities (flow velocity, buoyancy, abundance variance) scale with the luminosity of the model. This is similar to the work done by \cite{2019MNRAS.484.4645C} and \cite{Rizzuti2022b,Rizzuti2023b} for a carbon- and neon-burning shells respectively, however, for the first time, we span a large range of boosting comparing the nominal luminosity runs to those boosted by 1000. As it is not computationally feasible to run very low Mach number flows without boosting the luminosity \citep[e.g.][]{2017MNRAS.471..279C, 2021A&A...653A..55H},  this comparison is important as a first step to determine the feasibility and correctness of very high boosting. 

In general, increasing the boosting gave results which are in line with our previous findings \citep{2019MNRAS.484.4645C,Rizzuti2022b,Rizzuti2023b}. In particular, the convective velocities, rate of CBM and the corresponding growth of the convective shell all scale with the boosting applied. The present results (as well as those published in \citet{Rizzuti2022b} using the same simulations as the ones presented in this paper) show that these quantities for the non-boosted (nominal luminosity) case can be extrapolated from simulations with a higher boosting factor (typically 10x or 100x), which are cheaper to perform in terms of computing resources. The turbulent spectra in each case all show similar shapes, but with different energies, as expected. Inside the convective zone, we find a very efficient mixing, making the chemical composition in the bulk of the convective region extremely homogeneous for reasonable boosting factors.

For the highest boosting factor we used (\texttt{Ex1000} simulation), the nuclear burning becomes comparable to the transport timescale ($Da\approx 1$) in a significant fraction of the convection zone. This is qualitatively different to all previous boosting cases, where nuclear burning is only faster than the transport timescale in a very thin region at the base of the convective shell. In the \texttt{Ex1000}  case, the mixing is not efficient enough to homogenise the convective zone, and a chemical gradient becomes visible in the simulations. This leads to the composition variance showing significant deviations ($>10\%$) throughout the convection zone, breaking down the approximations that are normally used for 1D stellar evolution calculations. While in our study, this $Da\approx 1$  event occurs due to artificial luminosity boosting, they are expected to occur naturally in dynamical processes in stars, such as ingestion events \citep{Herwig2011a, Mocak2018a}, and possibly during shell mergers. 

Our results show that a correct treatment of convection in 1D stellar evolution codes is required to accurately follow the convective boundary mixing and reproduce the behaviour appearing in our 3D simulations. The assumption that the mixing is very efficient and almost instantaneous in convective regions appears to be valid, except in cases where the nuclear burning is very fast (Da$\sim 1$). Finally, we want to stress that the results presented here are valid for deep convection during advanced stages of stellar evolution, and we need to remain cautious when using them in other phases of the evolution, particularly for convection during the early stage (main sequence), or in stellar envelopes, where radiative effects play a key role.

\section*{Acknowledgements}
RH acknowledges support from the World Premier International Research Centre Initiative (WPI Initiative), MEXT, Japan and the IReNA AccelNet Network of Networks (National Science Foundation, Grant No. OISE-1927130). CG has received funding from the European Research Council (ERC) under the European Union’s Horizon 2020 research and innovation program (Grant No. 833925). WDA acknowledges support from the Theoretical Astrophysics Program (TAP) at the University of Arizona and Steward Observatory. CG, RH, and CM acknowledge ISSI, Bern, for its support in organising collaboration. This article is based upon work from the ChETEC COST Action (CA16117) and the European Union’s Horizon 2020 research and innovation programme (ChETEC-INFRA, Grant No. 101008324). The authors acknowledge PRACE for awarding access to the resource MareNostrum 4 at Barcelona Supercomputing Center, Spain, and the STFC DiRAC HPC Facility at Durham University, UK (Grants ST/P002293/1, ST/R002371/1, ST/R000832/1, ST/K00042X/1, ST/H008519/1, ST/ K00087X/1, ST/K003267/1). The University of Edinburgh is a charitable body, registered in Scotland, with Registration No. SC005336. For the purpose of open access, the author has applied a Creative Commons Attribution (CC BY) licence to any Author Accepted Manuscript version arising.

\section*{Data availability}
The data underlying this article will be shared on reasonable request to the corresponding author.

\bibliographystyle{mnras}
\bibliography{Biblio}

\end{document}